\documentclass{nature3}

\usepackage{graphicx}%
\usepackage{multirow}%
\usepackage{mhchem}
\usepackage{adjustbox}

\usepackage{titlesec}
\titleformat{\section}{\normalfont\large\bfseries}{\thesection.}{3pt}{\space}[]
\titlespacing*{\section}{0em}{1ex}{1em}[0em]
\titleformat{\subsection}{\bfseries}{\thesubsection.}{3pt}{\space}[]
\titlespacing*{\subsection}{0em}{1ex}{1em}[0em]
\usepackage{xcolor}
\usepackage[colorlinks,breaklinks=true,plainpages=false,citecolor=blue,linkcolor=blue,urlcolor=blue,bookmarksopen=true,bookmarksnumbered=false,bookmarksdepth=5]{hyperref}
\usepackage[symbol]{footmisc}
\renewcommand{\thefootnote}{\fnsymbol{footnote}}

\usepackage{afterpage}

\usepackage{comment}

\usepackage[normalem]{ulem}

\usepackage{rotating}

\usepackage{amsmath,amssymb,amsfonts}%
\usepackage{amsthm}%
\usepackage{mathrsfs}%
\usepackage[title]{appendix}%
\usepackage{xcolor}%
\usepackage{textcomp}%
\usepackage{manyfoot}%
\usepackage{booktabs}%
\usepackage{algorithm}%
\usepackage{algorithmicx}%
\usepackage{algpseudocode}%
\usepackage{listings}%
\usepackage{mhchem}

\usepackage{multibib}
\newcites{Met}{Methods References}
\newcount\longrefs
\def\aap{\ifnum\longrefs=1 {Astron.\ Astrophys.}\else 
                           {A\hbox{\rm \&}A}\fi}
\def\aapl{\ifnum\longrefs=1 {Astron.\ Astrophys.\ Lett.}\else 
                           {A\hbox{\rm \&}A}\fi}
\def\aapr{\ifnum\longrefs=1 {Astron.\ Astrophys.\ Rev.}\else 
                            {A\hbox{\rm \&}AR}\fi}
\def\aaps{\ifnum\longrefs=1 {Astron.\ Astrophys.\ Suppl.}\else 
                            {A\hbox{\rm \&}AS}\fi}
\def\aj{\ifnum\longrefs=1 {Astron.\ J.}\else 
                          {AJ}\fi} 
\def\ao{\ifnum\longrefs=1 {Applied Optics}\else 
                           {Appl.\ Opt.}\fi} 
\def\aspcs{\ifnum\longrefs=1 {Astron.\ Soc.\ Pacific Conf. Series}\else 
                           {ASP Conf.\ Ser.}\fi} 
\def\apj{\ifnum\longrefs=1 {Astrophys.\ J.}\else 
                           {ApJ}\fi} 
\def\apjl{\ifnum\longrefs=1 {Astrophys.\ J.\ Lett.}\else 
                            {ApJ}\fi} 
\def\aplett{\ifnum\longrefs=1 {Astrophys.\ J.\ Lett.}\else 
                            {ApJ}\fi} 
\def\apjs{\ifnum\longrefs=1 {Astrophys.\ J.\ Suppl.}\else 
                            {ApJS}\fi}
\def\apss{\ifnum\longrefs=1 {Astrophys.\ and Space Science}\else 
                            {Ap\hbox{\rm \&}SS}\fi}
\def\araa{\ifnum\longrefs=1 {Ann.\ Rev.\ Astron.\ Astrophys.}\else 
                            {ARA\hbox{\rm \&}A}\fi}
\def\azh{\ifnum\longrefs=1 {Astronomicheskii Zhurnal}\else 
                            {Astron.\ Zhur.}\fi}
\def\baas{\ifnum\longrefs=1 {Bull.\ Am.\ Astron.\ Soc.}\else 
                            {BAAS}\fi}
\def\bain{\ifnum\longrefs=1 {Bull.\ Astronom.\ Institutes Netherlands}\else
                            {Bull.\ Astr.\ Inst.\ Neth.}\fi}
\def\gca{\ifnum\longrefs=1 {Geochim.\ Cosmochim.\ Acta}\else 
                           {Geochim.\ Cosmochim.\ Acta}\fi}
\def\grl{\ifnum\longrefs=1 {Geophys.\ Res.\ Lett.}\else 
                           {Geoph.\ Res.\ Lett.}\fi}
\def\iaucirc{\ifnum\longrefs=1 {IAU Circulars}\else 
                          {IAU Circ.}\fi}
\def\ip{\ifnum\longrefs=1 {in press}\else 
                          {in press}\fi}
\def\jchemp{\ifnum\longrefs=1 {J.\ Chem.\ Phys.}\else 
                           {J.\ Chem.\ Phys.}\fi}  
\def\jcp{\ifnum\longrefs=1 {J.\ Chem.\ Phys.}\else 
                           {J.\ Chem.\ Phys.}\fi}  
\def\jgr{\ifnum\longrefs=1 {J.\ Geophys.\ Res.}\else 
                           {J.\ Geophys.\ Res.}\fi}  
\def\jmolspec{\ifnum\longrefs=1 {J.\ Mol.\ Spectrosc.}\else 
                           {J.\ Mol.\ Spectrosc.}\fi}  
\def\jqsrt{\ifnum\longrefs=1 {J.\ Quant.\ Spectrosc.\ Radiat.\ Transfer}\else 
                           {J.\ Quant.\ Spectrosc.\ Radiat.\ Transfer}\fi}  
\def\jrasc{\ifnum\longrefs=1 {J.\ Royal Astron.\ Soc.\ Canada}\else 
                           {JRAS Can.}\fi}  
\def\mnras{\ifnum\longrefs=1 {Mon.\ Not.\ Roy.\ Astron.\ Soc.}\else 
                             {MNRAS}\fi} 
\def\nat{\ifnum\longrefs=1 {Nature}\else 
                           {Nat}\fi}
\def\pasj{\ifnum\longrefs=1 {Pub.\ Astron.\ Soc.\ Japan}\else 
                            {PASJ}\fi} 
\def\pasp{\ifnum\longrefs=1 {Pub.\ Astron.\ Soc.\ Pacific}\else 
                            {PASP}\fi} 
\def\physscr{\ifnum\longrefs=1 {Physica Scripta}\else 
                            {Phys.\ Scrip.}\fi} 
\def\planss{\ifnum\longrefs=1 {Planetary \& Space Science}\else 
                            {Plan. \& Space Sci.}\fi} 
\def\procspie{\ifnum\longrefs=1 {Proc.\ SPIE}\else 
                            {Proc.\ SPIE}\fi} 
\def\qjras{\ifnum\longrefs=1 {Quarterly J.\ Royal Astron.\ Soc.}\else 
                            {QJRAS}\fi} 
\def\sa{\ifnum\longrefs=1 {Soviet Astron..}\else 
                               {Sov.\ Astron.}\fi}
\def\skytel{\ifnum\longrefs=1 {Sky \& Telescope}\else 
                            {Sky \& Tel.}\fi} 
\def\solphys{\ifnum\longrefs=1 {Solar Phys.}\else 
                               {Solar Phys.}\fi}
\def\ssr{\ifnum\longrefs=1 {Space Science Rev.}\else 
                               {Space\ Sci.\ Rev.}\fi}

\def\dutch{\def\refname{Referenties}\def\abstractname{Samenvatting}%
  \def\bibname{Bibliografie}\def\chaptername{Hoofdstuk}%
  \def\appendixname{Bijlage}\def\contentsname{Inhoudsopgave}%
  \def\listfigurename{Lijst van figuren}\def\listtablename{Lijst van tabellen}%
  \def\indexname{Index}\def\figurename{Figuur}\def\tablename{Tabel}%
  \def\partname{Deel}\def\enclname{Bijlage(n)}\def\ccname{Ter attentie van}%
  \def\headtoname{Aan}\def\headpagename{Pagina}%
  \def\today{\number\day\space\ifcase\month\or januari\or februari\or maart\or%
     april\or mei\or juni\or juli\or augustus\or september\or oktober\or%
     november\or december\fi \space\number\year}%
  \typeout{
              >>>>> use hlatex209 for Dutch hyphenation <<<<< 
         }}
\newcounter{onefig} \newcounter{fignumber}
\newcount\nocaptions \newcount\nofigures \newcount\figwidth
\newcount\viewgraphs
  \def\paper{}  \def\figlabel{} 
\long\def\nextfig#1{\setcounter{figure}{\value{fignumber}}
  \addtocounter{fignumber}{1}
  \ifnum \viewgraphs=1 \newpage \pagestyle{empty} \fi 
  \ifnum\value{onefig}=0 #1 \fi                 
  \ifnum\value{onefig}=\value{fignumber} #1 \fi}
\def\figwidths#1#2{\ifnum \nocaptions=1 #2mm \else #1mm \fi}  
\def\paper#1{}  
\long\def\plotfig#1#2{\ifnum \nofigures=1 \else #2 \fi}
\long\def\captiontext#1{\ifnum \nofigures=1 \raggedright \fi 
   \ifnum \nocaptions=1 \paper
     \ifnum \viewgraphs=0 
       \newline  \mbox{}\hrulefill\mbox{} \newline 
       \newline label:~\{\figlabel\} 
     \fi 
     \else \ifnum \nofigures=0 \fi 
   #1 \fi}

\newcount\panelwidth \newcount\panelheight 
\newcount\bxmin \newcount\bymin \newcount\bxmax \newcount\bymax
\newcount\tbxmin \newcount\tbymin
\newcount\tpanelwidth \newcount\tpanelheight \newcount\tpdif
\def\panelsize #1,#2;{\panelwidth=#1 \panelheight=#2}  
\def\setbb #1,#2;#3,#4;#5,#6;{
  \tbxmin=#1 \tbymin=#2    
  \bxmin=#3 \bymin=#4      
  \bxmax=#5 \bymax=#6}     
\def\barepanel #1{%
  \ifnum\panelheight=0 
    \tpdif=\bymax \advance\tpdif by -\bymin
    \multiply \tpdif by \panelwidth
    \tpanelheight=\tpdif
    \tpdif=\bxmax \advance\tpdif by -\bxmin
    \divide \tpanelheight by \tpdif
  \else \tpanelheight=\panelheight \fi
  \epsfig{file=#1,%
     bbllx=\bxmin bp,bblly=\bymin bp,bburx=\bxmax bp,bbury=\bymax bp,clip=,%
     width=\panelwidth mm,height=\tpanelheight mm}}
\def\labelypanel #1{
  \ifnum\panelheight=0 
    \tpdif=\bymax \advance\tpdif by -\bymin
    \multiply \tpdif by \panelwidth
    \tpanelheight=\tpdif
    \tpdif=\bxmax \advance\tpdif by -\bxmin
    \divide \tpanelheight by \tpdif
  \else \tpanelheight=\panelheight \fi
  \tpdif=\bxmax \advance\tpdif by -\tbxmin
  \tpanelwidth=\panelwidth \multiply \tpanelwidth by \tpdif
  \tpdif=\bxmax \advance\tpdif by -\bxmin
  \divide \tpanelwidth by \tpdif
  \epsfig{file=#1,%
    bbllx=\tbxmin bp,bblly=\bymin bp,bburx=\bxmax bp,bbury=\bymax bp,%
    clip=,width=\tpanelwidth mm,height=\tpanelheight mm}}
\def\labelxpanel #1{%
  \ifnum\panelheight=0 
    \tpdif=\bymax \advance\tpdif by -\bymin
    \multiply \tpdif by \panelwidth
    \tpanelheight=\tpdif
    \tpdif=\bxmax \advance\tpdif by -\bxmin
    \divide \tpanelheight by \tpdif
  \else \tpanelheight=\panelheight \fi
  \tpdif=\bymax \advance\tpdif by -\tbymin
  \multiply \tpanelheight by \tpdif
  \tpdif=\bymax \advance\tpdif by -\bymin
  \divide \tpanelheight by \tpdif
  \epsfig{file=#1,%
    bbllx=\bxmin bp,bblly=\tbymin bp,bburx=\bxmax bp,bbury=\bymax bp,%
    clip=,width=\panelwidth mm,height=\tpanelheight mm}}
\def\labelxypanel #1{%
  \ifnum\panelheight=0 
    \tpdif=\bymax \advance\tpdif by -\bymin
    \multiply \tpdif by \panelwidth
    \tpanelheight=\tpdif
    \tpdif=\bxmax \advance\tpdif by -\bxmin
    \divide \tpanelheight by \tpdif
  \else \tpanelheight=\panelheight \fi
  \tpdif=\bxmax \advance\tpdif by -\tbxmin
  \tpanelwidth=\panelwidth \multiply \tpanelwidth by \tpdif
  \tpdif=\bxmax \advance\tpdif by -\bxmin
  \divide \tpanelwidth by \tpdif 
  \tpdif=\bymax \advance\tpdif by -\tbymin 
  \multiply \tpanelheight by \tpdif
  \tpdif=\bymax \advance\tpdif by -\bymin
  \divide \tpanelheight by \tpdif
  \epsfig{file=#1,%
    bbllx=\tbxmin bp,bblly=\tbymin bp,bburx=\bxmax bp,bbury=\bymax bp,%
    clip=,width=\tpanelwidth mm,height=\tpanelheight mm}}

\def\CC{\par \vspace*{-2ex} \footnotesize \baselineskip=8pt \begin{verbatim}}

\long\def\startignore #1\stopignore{}   

\def\setlistparams{         
  \topsep=0.7ex                 
  \itemsep=0.7ex                
  \leftmargini=3ex}             
\setlistparams                  

\newcounter{alistindex}       

\def\spacing #1{\small \renewcommand{\baselinestretch}{#1}
                \normalsize \setlistparams}             

\newcounter{romenumnr}

\newlength{\minipagewidth}

\newsavebox{\boxcontent}
\newcommand{\ovalhead}[1]{
  \unitlength=1cm
  \sbox{\boxcontent}{\mbox{~~{#1}~~}}
  \begin{center}
    \ifdim\wd\boxcontent>6ex 
    \ifdim\wd\boxcontent<8cm 
    \begin{picture}(8,3) \thicklines     
      \put(4.0,0.8){\oval(8,1.6)} 
      \put(0.0,0.7){\parbox{8cm}{
         \begin{center} \usebox{\boxcontent} \end{center}}}
    \end{picture}
    \else \ifdim\wd\boxcontent<12cm 
    \begin{picture}(12,3) \thicklines     
        \put(6.0,0.8){\oval(12,1.6)} 
        \put(0.0,0.7){\parbox{12cm}{
           \begin{center} \usebox{\boxcontent} \end{center}}}
    \end{picture}
    \else
    \begin{picture}(16,3) \thicklines     
        \put(8.0,0.8){\oval(16,1.6)} 
        \put(0.0,0.7){\parbox{16cm}{
           \begin{center} \usebox{\boxcontent} \end{center}}}
    \end{picture}
    \fi \fi \fi
  \end{center}} 

\newcounter{headnr}            
\newcounter{subheadnr}[headnr]
\newcounter{subsubheadnr}[subheadnr]
\def\head #1\par{
  \stepcounter{headnr}                          
  \vspace{2ex} \noindent                        
  {\bf \theheadnr~~~~#1}\\[1ex] \noindent}      
\def\subhead #1\par{  
  \stepcounter{subheadnr}
  \vspace{1.3ex} \noindent
  {\bf \theheadnr.\arabic{subheadnr}~~~#1}\\[0.3ex] \noindent}
\def\subsubhead #1\par{
  \stepcounter{subsubheadnr}
  \vspace{1.0ex} \noindent
  {\bf \theheadnr.\arabic{subheadnr}.\arabic{subsubheadnr}~~~#1}\\ \noindent}

\font\dropfont= cmr12 scaled \magstep5
\def\dropcap#1#2{{\noindent
    \setbox0\hbox{\dropfont #1}\setbox1\hbox{#2}\setbox2\hbox{(}%
    \count0=\ht0\advance\count0 by\dp0\count1\baselineskip
    \advance\count0 by-\ht1\advance\count0by\ht2
    \dimen1=.5ex\advance\count0by\dimen1\divide\count0 by\count1
    \advance\count0 by1\dimen0\wd0
    \advance\dimen0 by.25em\dimen1=\ht0\advance\dimen1 by-\ht1
    \global\hangindent\dimen0\global\hangafter-\count0
    \hskip-\dimen0\setbox0\hbox to\dimen0{\raise-\dimen1\box0\hss}%
    \dp0=0in\ht0=0in\box0}#2}

\def\level #1 #2#3#4{$#1 \: ^{#2} \mbox{#3} ^{#4}$}   

\def\mathstacksym#1#2#3#4#5{\def#1{\mathrel{\hbox to 0pt{\lower 
    #5\hbox{#3}\hss} \raise #4\hbox{#2}}}}

\mathstacksym\lta{$<$}{$\sim$}{1.5pt}{3.5pt} 
\mathstacksym\gta{$>$}{$\sim$}{1.5pt}{3.5pt} 
\mathstacksym\lrarrow{$\leftarrow$}{$\rightarrow$}{2pt}{1pt} 
\mathstacksym\lessgreat{$>$}{$<$}{3pt}{3pt} 

\title{$^{15}$NH$_3$ in the atmosphere of a cool brown dwarf
}

\begin{document}

\renewcommand{\thefootnote}{\alph{footnote}}

\maketitle

\noindent\author{
David Barrado$^{1,\star,\dagger}$,
Paul Molli\`ere$^{2,\dagger}$,
Polychronis Patapis$^{3,\dagger}$,
Michiel Min$^{4}$, 
Pascal Tremblin$^{5}$, 
Francisco Ardevol Martinez$^{6,7,8}$, 
Niall Whiteford$^{9}$, 
Malavika Vasist$^{10}$, 
Ioannis Argyriou$^{11}$,
Matthias Samland$^{2}$, 
Pierre-Olivier Lagage$^{12}$, 
Leen Decin$^{11}$, 
Rens Waters$^{13,14}$, 
Thomas Henning$^{2}$, 
Mar\'{\i}a Morales-Calder\'on$^{1}$, 
Manuel Guedel$^{15,2,16}$, 
Bart Vandenbussche$^{11}$, 
Olivier Absil$^{10}$, 
Pierre Baudoz$^{17}$, 
Anthony Boccaletti$^{17}$, 
Jeroen Bouwman$^{2}$, 
Christophe Cossou$^{18}$, 
Alain Coulais$^{19,12}$, 
Nicolas Crouzet$^{20}$,
Ren\'e Gastaud$^{21}$, 
Alistair Glasse$^{22}$, 
Adrian M. Glauser$^{3}$, 
Inga Kamp$^{6}$, 
Sarah Kendrew$^{23}$, 
Oliver Krause$^{2}$, 
Fred Lahuis$^{4}$, 
Michael Mueller$^{6}$, 
G\"oran Olofsson$^{24}$, 
John Pye$^{25}$, 
Daniel Rouan$^{17}$, 
Pierre Royer$^{11}$, 
Silvia Scheithauer$^{2}$, 
Ingo Waldmann$^{26}$,
Luis Colina$^{1}$, 
Ewine F. van Dishoeck$^{20}$, 
Tom Ray$^{27}$, 
G\"oran \"Ostlin$^{28}$, 
Gillian Wright$^{29}$
}

\noindent$^\dagger$These authors contributed equally to this work. \\
$^\star$Corresponding author(s). Email(s): barrado@cab.inta-csic.es.\\

\begin{affiliations}
\item 
Centro de Astrobiolog\'{\i}a (CAB), CSIC-INTA, ESAC Campus, Camino Bajo del Castillo s/n, 28692 Villanueva de la
Ca\~nada, Madrid, Spain
 \item 
 Max-Planck-Institut f\"ur Astronomie (MPIA), K\"onigstuhl 17, 69117 Heidelberg, Germany
\item
Institute of Particle Physics and Astrophysics,  ETH Z\"urich, Wolfgang-Pauli-Strasse 27, 8093, Z\"urich, Switzerland
\item
SRON Netherlands Institute for Space Research, Niels Bohrweg 4, 2333 CA Leiden, The Netherlands
\item
Universit\'e Paris-Saclay, UVSQ, CNRS, CEA, Maison de la Simulation, 91191, Gif-sur-Yvette, France
\item
Kapteyn Institute of Astronomy, University of Groningen, Groningen, The Netherlands
\item
Netherlands Institute for Space Research (SRON), Leiden, The Netherlands
\item
School of GeoSciences and Centre for Exoplanet Science, University of Edinburgh, Edinburgh, United Kingdom
\item
Department of Astrophysics, American Museum of Natural History, New York, NY 10024, USA
\item
STAR Institute, Universit\'e de Li\`ege, All\'ee du Six Ao\^ut 19c, 4000 Li\`ege, Belgium
\item
Institute of Astronomy, KU Leuven, Celestijnenlaan 200D, 3001 Leuven, Belgium
\item
Universit\'e Paris-Saclay, Universit\'e Paris cit\'e, CEA, CNRS, AIM,  91191, Gif-sur-Yvette cedex, France
\item
Department of Astrophysics/IMAPP, Radboud University, PO Box 9010, 6500 GL Nijmegen, The Netherlands
\item 
SRON Netherlands Institute for Space Research, Niels Bohrweg 4, NL-2333 CA Leiden, the Netherlands
\item
Department of Astrophysics, University of Vienna, T\"urkenschanzstr. 17, 1180 Vienna, Austria
\item
ETH Z\"urich, Institute for Particle Physics and Astrophysics, Wolfgang-Pauli-Str. 27, 8093 Z\"urich, Switzerland
\item
LESIA, Observatoire de Paris, Universit\'e PSL, CNRS, Sorbonne Universit\'e, Universit\'e de Paris Cit\'e, 5 place Jules Janssen, 92195 Meudon, France
\item
Universit\'e Paris-Saclay, CEA, D\'epartement d'Electronique des D\'etecteurs et d'Informatique pour la Physique, 91191, Gif-sur-Yvette, France
\item
LERMA, Observatoire de Paris, Universit\'e PSL, CNRS, Sorbonne Universit\'e, Paris, France 
\item
Leiden Observatory, Leiden University, P.O. Box 9513, 2300 RA Leiden, The Netherlands
\item
Universit\'e Paris-Saclay, Universit\'e Paris Cit\'e, CEA, CNRS, AIM, F-91191 Gif-sur-Yvette, France
\item
UK Astronomy Technology Centre, Royal Observatory, Blackford Hill, Edinburgh EH9 3HJ, United Kingdom
\item
European Space Agency, Space Telescope Science Institute, Baltimore, MD, USA
\item
Department of Astronomy, Stockholm University, AlbaNova University Center, 10691 Stockholm, Sweden
\item
School of Physics \& Astronomy, Space Research Centre, Space Park Leicester, University of Leicester, 92 Corporation Road, Leicester, LE4 5SP, United Kingdom
\item
Department of Physics and Astronomy, University College London, Gower Street, WC1E 6BT, United Kingdom
\item
School of Cosmic Physics, Dublin Institute for Advanced Studies, 31 Fitzwilliam Place, Dublin, D02 XF86, Ireland
\item
Department of Astronomy, Oskar Klein Centre, Stockholm University, 106 91 Stockholm, Sweden
\item
UK Astronomy Technology Centre, Royal Observatory Edinburgh, Blackford Hill, Edinburgh EH9 3HJ, United Kingdom

\end{affiliations}

\maketitle

\setlength{\parskip}{11pt}

\bigskip
%
\begin{abstract} 
    Brown dwarfs serve as ideal laboratories for studying the atmospheres of giant exoplanets on wide orbits as the governing physical and chemical processes in them are nearly identical\cite{burrows1997nongray,faherty2018spectral}. Understanding the formation of gas giant planets is challenging, often involving the endeavour to link atmospheric abundance ratios, such as the carbon-to-oxygen (C/O) ratio, to formation scenarios\cite{Madhusudhan2014migration}. However, the complexity of planet formation requires additional tracers, as the unambiguous interpretation of the measured C/O ratio is fraught with complexity\cite{Molliere2022assumptions}.
    Isotope ratios, such as deuterium-to-hydrogen and $^{14}$N/$^{15}$N, offer a promising avenue to gain further insight into this formation process, mirroring their utility within the solar system\cite{Feuchtgruber2013_D_H_UranusNeptune}$^,$\cite{Alibert2018NatAs...2..873A}$^,$\cite{nomura2022isotopic}. For exoplanets only a handful of constraints on $^{12}$C/$^{13}$C exist, pointing to the accretion of $^{13}$C-rich ice from beyond the disks' CO iceline\cite{zhang2021a,Line2021C_O_HotAtmos}. Here we report on the mid-infrared detection of the $^{14}$NH$_3$ and $^{15}$NH$_3$ isotopologues in the atmosphere of a cool brown dwarf with an effective temperature of 380 K in a spectrum taken with the Mid-InfraRed Instrument of the James Webb Space Telescope.  As expected, our results reveal a $^{14}$N/$^{15}$N value consistent with star-like formation by gravitational collapse, demonstrating that this ratio can be accurately constrained. Since young stars and their planets should be more strongly enriched in the $^{15}$N isotope\cite{adande2012millimeter}, we expect that $^{15}$NH$_3$ will be detectable in a number of cold, wide-separation exoplanets.
\end{abstract}

\bigskip


The coldest class of brown dwarfs, so-called Y-dwarfs, span temperatures from 250~K to 500~K\cite{cushing2011discovery}. Their atmospheres are dominated by the absorption of water, methane, and ammonia, while water clouds likely become important for the colder Y-dwarfs\cite{Morley2018_Lspectrum_Ydwarf}. Since their emission peaks in the mid-infrared beyond 4~$\mu$m, Y-dwarf spectroscopic characterisation is challenging and the number of studies has been limited\cite{Skemer2016_Spectra_Ydwarfs}$^,$\cite{Morley2018_Lspectrum_Ydwarf}$^,$\cite{Cushing2021_J1828}. JWST is transforming the study of Y-dwarfs, by granting access to their full luminous range\cite{beilercushing2023}. We analysed JWST/MIRI Medium Resolution Spectrometer (MRS)\cite{Argyriou2023_MRS} observations of the Y-dwarf archetype WISEP J182831.08$+$265037.8 (hereafter WISE J1828), with an effective temperature of $\sim$380~K\cite{cushing2011discovery}. We obtain a mid-infrared spectrum at R$\sim$3,000 to 1,500, between 4.9 and 27.9~$\mu$m. The data reduction is described in Methods. 
Our observations are presented in Fig.~1,
together with an exemplary best-fit model from our analysis, and reveal a spectrum rich in molecular features, namely a broad water absorption band at 5-7~$\mu$m, methane at 7.6~$\mu$m, and ammonia at 9-13~$\mu$m. The ammonia band is also shown in more detail in the lower panels of Fig.~1.
We analyzed the atmospheric properties of WISE~J1828 using several retrieval codes\cite{Burningham2017_Retrieval}$^,$\cite{molliere_wardenier_2019}$^,$\cite{min_2020} and self-consistent atmosphere models in radiative-convective equilibrium\cite{Tremblin2015Fingering}$^,$\cite{chubbmin2022}$^,$\cite{ardevol23}. The best-fit spectra and residuals are shown in Extended Data, Fig.~1.
Since the MIRI observations mostly probe high altitudes in the atmosphere (with the contribution function peaking at $\sim$1 bar), we also added archival near-infrared data\cite{Cushing2021_J1828}, which probes deeper layers, at pressures of $\sim$10 bar (Extended Data Fig.~2).
Due to its high luminosity, given its spectral type, WISE J1828 is suggested to be a binary system\footnote{However, numerous studies, including recent JWST measurements, all failed at resolving its binarity, putting an upper limit of 0.5 au on the separation of its components\cite{cushing2011discovery,defurio2023jwst}.}. We thus modeled WISE~J1828 as an equal mass binary system, emitting with identical atmospheres.

By combining the results of different retrieval approaches (see Methods), we constrain the ${\rm log}_{10}($Volume  Mixing  Ratios$)$ (VMRs) of the conspicuous absorbers H$_2$O, CH$_4$, and $^{14}$NH$_3$ to be $-3.03_{-0.21}^{+0.18}$, $-3.65_{-0.21}^{+0.21}$, and $-4.79_{-0.25}^{+0.15}$ respectively. The surface gravity of WISE~J1828 is constrained to be ${\rm log}(g)$=$4.34_{-0.88}^{+0.42}$, the effective temperature is $378_{-18}^{+13}$~K, and the radius is constrained to $1.37_{-0.13}^{+0.26} \ {\rm R_{Jup}}$. The uncertainties for these values are dominated by the dispersion between the various fitting approaches, and are thus larger than the actual uncertainties derived from any single analysis. A posterior plot of the various retrievals is shown in the Extended Data Fig.~3,
while all retrieval results are summarized in the Extended Data Table~1.
The self-consistent models constrain the atmospheric properties to be ${\rm log}(g)=4.5\pm1.0$, $R=1.27\pm0.21 \ {\rm R_{Jup}}$, and $T_{\rm eff}=450\pm101$~K. Again, the reported uncertainties are dominated by differences between the two models.
Our best-fit values of ${\rm log}(g)$, $R$, and $T_{\rm eff}$, after applying a binary correction to the inferred radii (dividing by $\sqrt{2}$), indicate an age of $\sim$5~Gyr for a $\sim$15 $\rm M_{Jup}$ equal-mass binary system\cite{Phillips2020_Models_TYdwarfs}, which is consistent with our mass constraints, see Extended Data Table~2.

The metallicity we derive for WISE~J1828, combining the results of all approaches that included it as a free parameter (retrievals and self-consistent), is consistent with solar: ${\rm [M/H]}=0.02_{-0.31}^{+0.12}$. For C/O we find $0.22_{-0.03}^{+0.37}$ (solar is 0.55$\pm$0.10\cite{Asplund2009_Sun_COmposition}), while N/O is constrained to $0.014_{-0.002}^{+0.021}$ (solar is 0.138$\pm$0.023\cite{Asplund2009_Sun_COmposition}). These uncertainties are again dominated by the dispersion between different approaches. The resulting posteriors for the metallicity, C/O and N/O are shown in Extended Data Fig.~3.
Our findings thus indicate an atmosphere with a solar bulk metallicity, but depleted in C and N. A likely cause for this is a departure from chemical equilibrium, where gas poor in both NH$_3$ and CH$_4$ is mixed up from the deep interior of the object\cite{Zahnle2014_CH4_CO_NH3_BD}.  The resulting gas would be enriched in \ce{N2} and \ce{CO}, which our observations are not sensitive to; while \ce{N2} is not spectrally active at all, our shortest wavelengths are longer than the location of the fundamental band of \ce{CO}  at $\sim$4.5~$\mu$m. However, it is questionable whether enough CO can be mixed up from the deep atmosphere to palpably change the inferred C/O ratio\cite{milesskemer2020}.

In addition, we detect $^{15}$NH$_3$ with a significance ranging from 4-6~$\sigma$, with several lines of ammonia clearly visible in the data, see Fig.~1.
We derive a VMR of $-7.68_{-0.34}^{+0.24}$ for $^{15}$NH$_3$ and $^{14}$N/$^{15}$N=$670_{-211}^{+390}$, averaging over the results of various models. In Fig.~2,
we summarize $^{14}$N/$^{15}$N for a range of astrophysical objects\cite{nomura2022isotopic}. Our value for WISE J1828 is consistent with solar at the 1-2~$\sigma$ level. Both the Sun and WISE~J1828, which we derive to have similar ages, have $^{14}$N/$^{15}$N values above those observed in the interstellar medium, which has been enriched in $^{15}$N by galactic stellar evolution since their formation\cite{adande2012millimeter}. Our measurement thus shows that WISE~J1828 most likely formed like a star, as expected\cite{chabrier2014giant}. A strong ice enrichment is unlikely, and we correspondingly rule out cometary values $^{14}$N/$^{15}$N$<$200 by more than 3 $\sigma$.

Constraints on $^{14}$N/$^{15}$N can serve as a formation tracer. For example, comets in the solar system, fundamental planetary building blocks, are enriched in $^{15}$N by a factor of 2-3 when compared to solar, due to $^{15}$N-rich NH$_3$ and HCN ice. In contrast, N$_2$ gas in the solar accretion disk is thought to have been depleted in $^{15}$N\cite{nomura2022isotopic}. In the solar system both Jupiter and Saturn are enriched in bulk nitrogen, but show $^{14}$N/$^{15}$N values consistent with the Sun. This may mean that they accreted ice cold enough to contain even the volatile N$_2$ ice, which requires temperatures $<$30~K, and corresponds to orbital distances $>$25~au\cite{fletcher2014origin,Oberg2019_Jupiter_Snowline}. An enriched nitrogen content through accretion of ice, but at solar $^{14}$N/$^{15}$N, may therefore require accretion close to where even the highly volatile N$_2$ can condense.

The understanding of how nitrogen fractionation actually occurs is incomplete\cite{nomura2022isotopic}, but we summarize some processes during different stages of the stellar and planetary evolution in Fig.~3.
In the denser parts of the clouds an increase of $^{15}$N in NH$_3$ is inferred, with a candidate for this fractionation being isotope-selective photo-dissociation\cite{furuya2018depletion}. Subsequently, NH$_3$ and HCN ices condense in the colder clumps, potentially producing $^{15}$N-rich ice. Once a protostar is formed, there is some evidence that $^{14}$N/$^{15}$N decreases further\cite{bergner2020evolutionary,Visser2018_N_disks}. The final consequence is thought to be a $^{15}$N increase in the less volatile nitrogen-carriers, leading to the observed increase in $^{15}$N in HCN  for protoplanetary disks and in \ce{NH3} and HCN for solar system comets. We note that models predict $^{15}$N-poor \ce{NH3} gas in protoplanetary disks\cite{Visser2018_N_disks}.

To further assess $^{14}$N/$^{15}$N as a formation tracer, we used a simplified planet formation model\cite{Molliere2022assumptions}. We tracked $^{14}$N/$^{15}$N as a function of the mass accreted as icy and rocky material, for a planet located between the N$_2$ iceline of the disk, at about 20-80~au, and the NH$_3$ iceline, ten times closer in\cite{Oberg2019_Jupiter_Snowline}$^,$\cite{bosman2019jupiter}. The ices were therefore likely enriched in $^{15}$N. We find that for Saturn-like metal enrichment ($\sim$6$\times$solar\cite{guillot2015}) the $^{14}$N/$^{15}$N decreases by 30-40\% when compared to solar (see Extended Data Fig.~4),
indicating that $^{14}$N/$^{15}$N can significantly vary when compared to the stellar value for a planet forming between the N$_2$ and NH$_3$ icelines.

With JWST MIRI, the formation-sensitive isotopologues $^{14}$NH$_3$ and $^{15}$NH$_3$ become accessible  for objects with low effective temperatures. In the mid-infrared, \ce{NH3} is a dominating absorber from $T_{\rm eff}=1000$~K\cite{suarez2022ultracool}, down to at least 380~K, the effective temperature of WISE~J1828. As demonstrated above, $^{14}$N/$^{15}$N can constrain formation locations with respect to the NH$_3$ and N$_2$ icelines of the disk. This is in addition to constraints on N/O, that the detection of H$_2$O and NH$_3$ enables, which has been suggested as another useful formation tracer\cite{Oberg_Bergin2021, turrini_2021}, but which needs careful interpretation due to chemical disequilibrium processes\cite{Zahnle2014_CH4_CO_NH3_BD}. Simultaneous constraints on C/O, N/O and $^{14}$N/$^{15}$N, {\rm based on \ce{CH4}, \ce{CO}, \ce{H2O} and \ce{NH3},} can be obtained for cold directly imaged exoplanets, further elucidating their formation history. These planets are found in orbits ranging from ten to hundreds of au, challenging the core-accretion paradigm for planetary formation\cite{Adams2021_CoreAccretion}. They either formed at their detected locations via a star-like gravitational instability, or originated closer to their star via core accretion and subsequent outward migration\cite{Marleau2019_Coreaccetion_HIP65426b}.



\section*{References}





  \begin{figure*}[!ht]
\centering
\includegraphics[width=\textwidth]{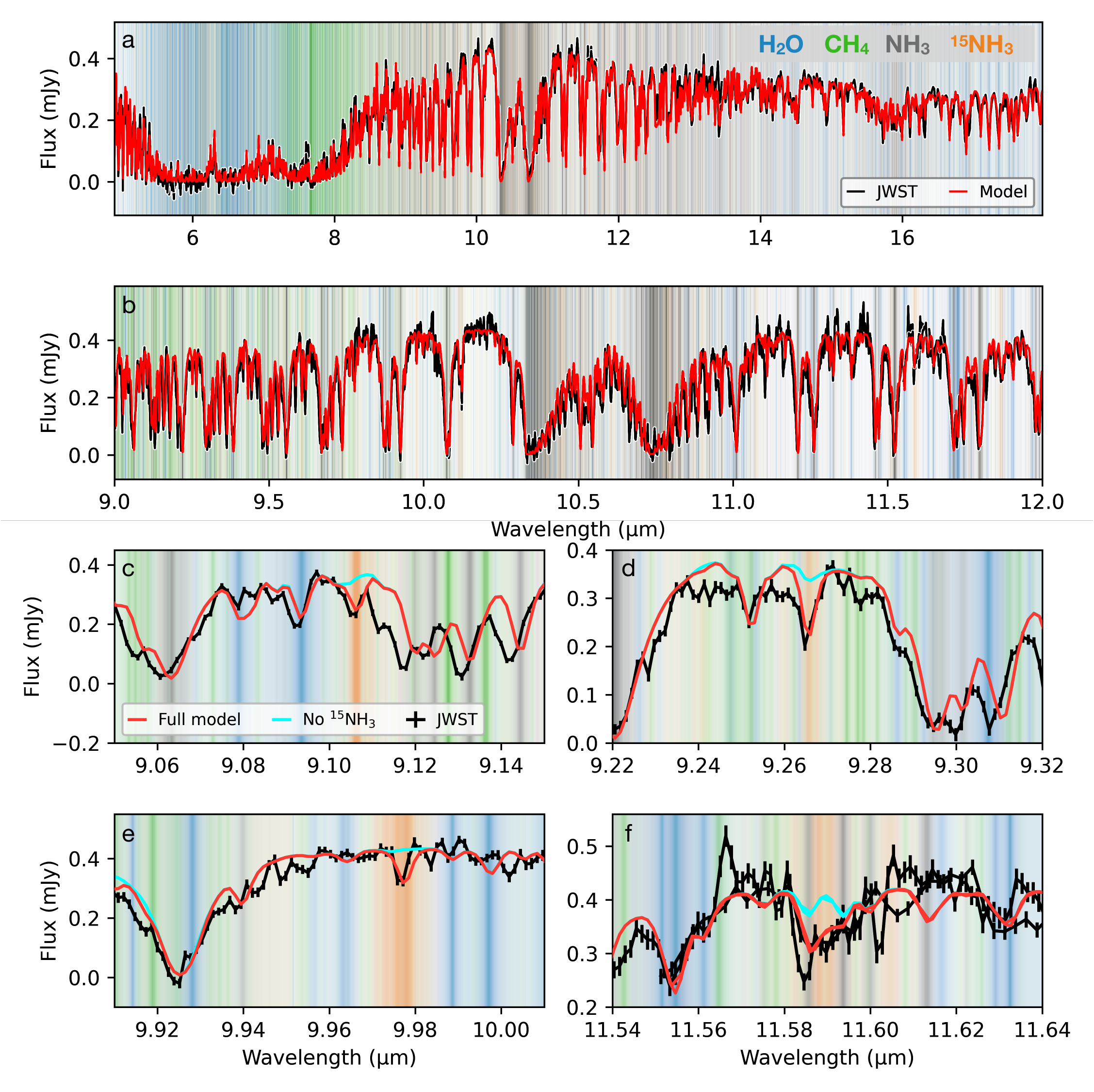}
\caption{\textbf{MIRI/MRS spectrum  and exemplary best-fit model (here: pRT-free) of the Y-dwarf WISE~J1828}. Panel a: full MIRI wavelength range considered in our models at retrieval resolution ($\lambda/\Delta \lambda = 1000$). All other panels show the data at the original, higher MRS resolution; models have been post-processed to the same resolution. Panel b: like a, but zoomed in to to show the \ce{NH3} absorption band at 10 $\mu$m in more detail. Panels c to f: individual $^{15}$NH$_3$ lines in the data, including a best-fit retrieval model with and without accounting for the opacity of $^{15}$NH$_3$. The error bars shown for the observations correspond to 1-$\sigma$ confidence levels. Panel f contains two overlapping MIRI MRS sub-channels. Colored lines denote the theoretical positions of the absorption lines of H$_2$O, CH$_4$, NH$_3$ and $^{15}$NH$_3$.}
\end{figure*}

\begin{figure*}[!ht]
\centering
\includegraphics[width=\textwidth]{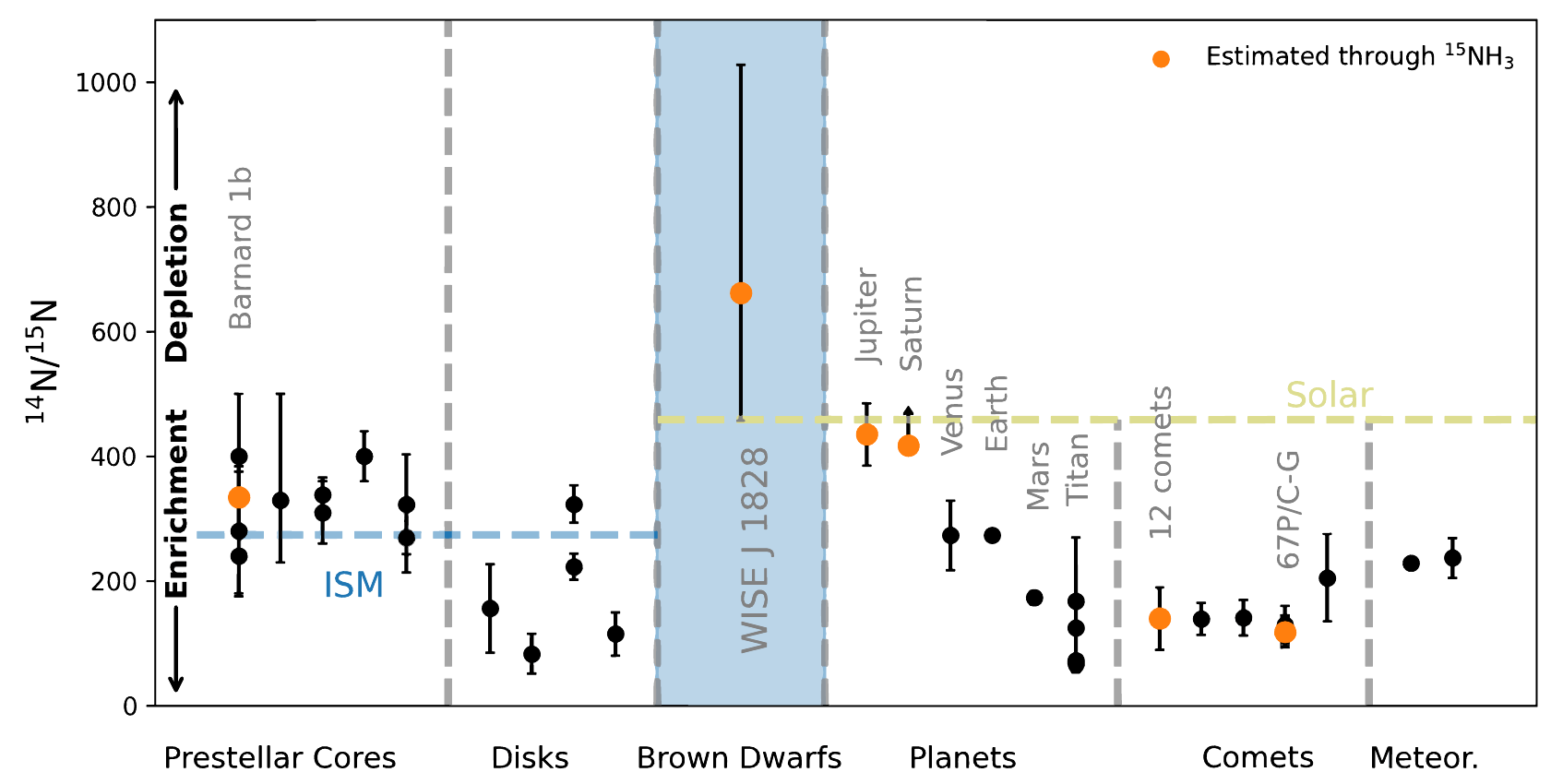}
\caption{\textbf{Comparison of the $^{14}$N/$^{15}$N ratio in the solar neighbourhood.} The values are based on either ammonia isotopologues (orange circles) or other molecules (black symbols), in different environments, subdivided by classes\cite{nomura2022isotopic}. Our estimate for WISE~J1828 appears as the brown dwarf class and is consistent with solar values (dashed horizontal yellow line)  at the 1-2~$\sigma$ level. Lower values indicate enrichment in $^{15}$N, while the dashed blue line represents the current value of the interstellar medium (ISM).}
\end{figure*}

\begin{figure*}[!ht]
\centering
\includegraphics[width=0.75\textwidth]{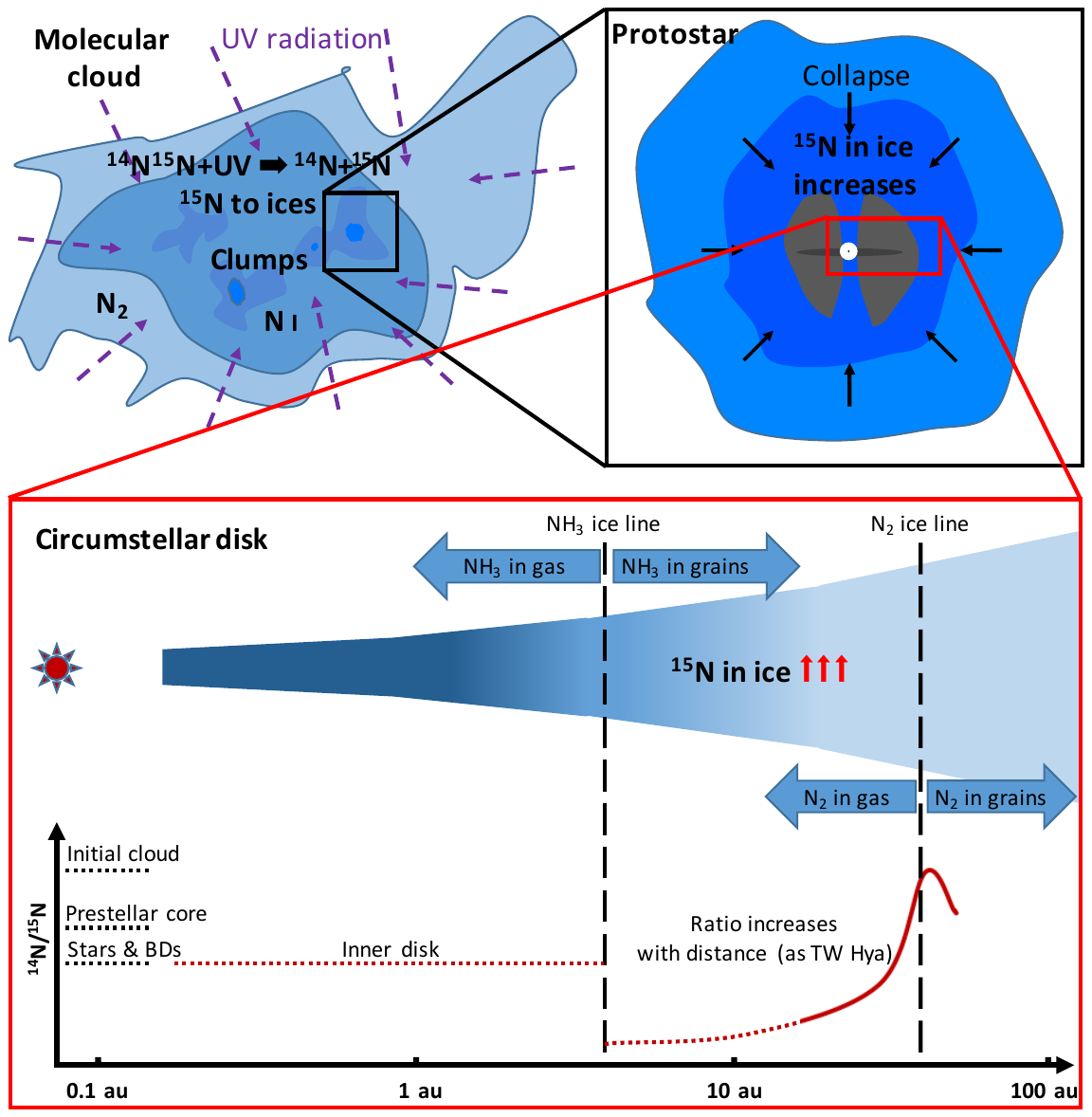}
\caption{\textbf{Different phases of the star and planetary formation}. We also show the relationship with the ammonia fractionation and the evolution of the $^{14}$N/$^{15}$N ratio at these stages: inside a molecular clouds with prestellar cores, during the formation of a protostar and in a circunstellar disk around a young star.}
\end{figure*}

\newpage


\begin{methods}\label{sec:methods}

\section*{JWST/MIRI observations and data processing}\label{Methods:JWST_data_reduction}

The MIRI/MRS targeted WISE~J1828 on July 28th 2022 without the use of target acquisition as part of JWST Guaranteed Time Observing program with PID 1189. All three dichroic/grating settings (SHORT, MEDIUM, LONG)  were obtained in order to cover the full wavelength range, from 4.9 - 27.9 $\mu$m\cite{Argyriou2023_MRS}. The observations were executed with the point source optimized two-point dither (negative direction), and the detector set up with 180 frames per integration. No dedicated background observations were obtained.

The jwst pipeline was used (pipeline version: 1.9, CRDS version: 11.16.20, CRDS context: jwst\_1045.pmap) to process the data. The raw (level 1B) files were processed with the detector level pipeline (CALWEBB\_DETECTOR1) to produce calibrated rate files. This pipeline applies corrections for the  nonlinearity of the ramps, dark current, detects jumps in the ramp due to cosmic rays, and finally fits the ramp signal to obtain slope values (rate.fits). The detector images were inspected and no sign of cosmic ray showers were found\cite{Argyriou2023_MRS}. Since the WISE~J1828 detector images were clean, we could use the fact that the point source itself is faint to perform a nod subtraction between the two dither points. First, we had to make sure that for every pair of detector images the flux levels were on the same level before subtracting them. A small time dependent difference between the individual integrations has been seen in MRS observations, with the first integration of the visit having a brighter flux level originating from the detector idling prior to the start of the exposures\cite{Argyriou2023_MRS}. We used the region between the MRS channels on the detectors that do not contain any astrophysical signal to estimate a single value using the median, and subtract it from the whole detector. Next, the two dithers were subtracted from each other to remove the thermal background contribution. For Channel 1 of the MRS, a detector artifact that manifests as vertical stripes was still present after the nod subtraction. These stripes were around 10\% of the science signal, and since the dispersion direction also closely follows the detector columns\cite{Argyriou2023_MRS} the stripes affect the continuum of the extracted spectrum. We chose a region of the detector where the source signal was almost zero, and estimated the stripe contribution as the median of ten rows, which was then subtracted from every row of the detector.

With the detector images clean from detector artifacts and the thermal background, the spectroscopy pipeline (CALWEBB\_SPEC2) was run in order to obtain calibrated detector images (cal.fits), assigning the astrometric and wavelength information, correcting for the scattered light and detector fringing, and applying the photometric calibration. Finally, with CALWEBB\_SPEC3 we built the dither combined cubes\cite{Law2023}, with outlier rejection enabled. With background subtracted cubes, the spectral extraction from the cube is done by performing aperture photometry for each wavelength slice of the cube. We first determined the center of the point source by fitting a 2D Gaussian in the wavelength collapsed cube, then placed an aperture of 1 Full Width Half Max (FWHM) centered on the source, applying an aperture correction for each wavelength to account for the missing flux of the point spread function outside the aperture. Some outliers remain in the extracted spectrum, which were removed for plotting the spectrum in Fig.~1
but not while fitting it. We traced back these outliers to the detector where a few cosmic rays overlap with the spectral trace, but are not bright enough for the outlier algorithm to detect. We therefore clip these values manually for each spectral band by setting a threshold for the flux. The outliers affected in total 0.4\% of the spectrum.

\section*{Retrieval analysis}\label{method:retrieval}
We carried out independent analyses of the MIRI/MRS spectra using a diversity of models, namely with the radiative-convective equilibrium codes ATMO and ARCiS+ML, and with the retrieval codes ARCiS, Brewster, and petitRADTRANS\cite{Tremblin2015Fingering}$^,$\cite{ min_2020}$^,$\cite{ Burningham2017_Retrieval}$^,$\cite{ molliere_wardenier_2019, chubbmin2022,ardevol23}. For computational feasibility, the retrievals with ARCiS, Brewster, and petitRADTRANS were run at a resolution of $\lambda/\Delta \lambda = 1000$; the MRS data was binned down correspondingly. For the retrievals we decided on a setup that assumed vertically constant absorber abundances that were retrieved freely, and a flexible parameterization of the pressure-temperature structure, which varied slightly between the setups, see below. For the retrievals presented here we assumed WISE~J1828 to be a single object, but allowed for a radius prior wide enough to account for an equal-brightness binary scenario. Clouds were neglected, which appears to be justified from a population-wide Y-dwarf retrieval analysis on Hubble Space Telescope (HST) data\cite{Zalesky2019_RetrievalYdwarfs}. We note, however, that the impact of clouds should increase towards longer wavelenghts, and for colder Y-dwarfs\cite{morleyskemer2018,manggao2022}. As shown in Extended Data Fig.~2,
our inferred $P$-$T$ profiles cross the water saturation vapor pressure curve at the top of the photosphere probed by HST and MIRI, so a cloud could have some moderate impact on our results and the effect of its inclusion should be assessed in future studies.

We also observed that the reported uncertainties of the JWST reduction can be much smaller than the differences observed in the overlapping regions of MRS subchannels. In addition, all best-fit models had a $\chi^2$ considerably larger than the number of wavelength channels. We thus opted for retrieving the magnitude of the uncertainties via the $10^b$ treatment\cite{lineteske2015}, where the error bars $\sigma$ considered during the retrieval are calculated from the uncertainties reported from the reduction $\sigma_{\rm red}$ as follows: $\sigma = \sqrt{\sigma_{\rm red}^2+10^b}$. Separate $b$s were retrieved for MIRI and HST data, respectively. The retrieved pressure-temperature structures of the retrievals would also exhibit kinks sometimes, which are challenging to reconcile with radiative-convective equilibrium models. In this case we implemented a regularization of the pressure-temperature structure\cite{lineteske2015}. With this modification, the ARCiS and petitRADTRANS retrievals optionally put a penalty on $d^2{\rm log}T/d{\rm log}P^2$, which strives towards a constant power-law dependence between pressure and temperature, since $d{\rm log}T/d{\rm log}P={\rm cst}$ implies $T\propto P^\alpha$, with $\alpha$ being the constant power law coefficient. This setup may therefore also reproduce the relation between pressure and temperature in the deep atmosphere, which is expected to be convective. The individual models we used for the analysis are described below.

The results of all model inferences for WISE~J1828 are found in Extended Data Table~1,
and the combined result of all retrievals is presented in Extended Data Table~2.
Elemental abundance ratios (C/O, N/O, $^{14}$N/$^{15}$N) were computed from the retrievals' VMR constraints for the atmospheric absorbers. They may thus miss additional atoms locked up in clouds (in the case of oxygen) or affected by quenching in species which are spectrally inactive (N$_2$ in the case of N) or have features outside the HST and MIRI wavelength range (CO in the case of C and O). The one-dimensional projection of the posteriors, for key forward model parameters, is shown in Extended Data Fig.~3
for all individual models. In Extended Data Fig.~2
we show the associated pressure-temperature uncertainties derived from all model analyses.

\subsection*{ATMO}
We briefly describe the main properties of the ATMO\cite{Tremblin2015Fingering} models and the grids that have been used in our study. These grids are publicly available at \url{https://opendata.erc-atmo.eu}. The ATMO models assume that clouds are not needed to reproduce the shape of the SED of brown dwarfs (apart from the 10-$\mu$m silicate feature). The authors have proposed that diabatic convective processes\cite{Tremblin2019} induced by out-of-equilibrium chemistry of CO/\ce{CH4} and \ce{N2}/\ce{NH3} can reduce the temperature gradient in the atmosphere and reproduce the reddening previously thought to occur by clouds. The grids assume a modification of the temperature gradient with an effective adiabatic index. The levels modified are between 2 and 200 bars at ${\rm log}(g)=5.0$ and are scaled by $10^{{\rm log}(g)-5}$ at other surface gravities. Out-of-equilibrium chemistry is used with $K_{zz} = 10^5$ cm$^2$ s$^{-1}$ at ${\rm log}(g) = 5.0$ and is scaled by $10^{2[5-{\rm log}(g)]}$ at other surface gravities. The mixing length is assumed to be two scale heights at 20 bars and higher pressures at ${\rm log}(g) = 5.0$, and is scaled down by the ratio between the local pressure and the pressure at 20 bars for lower pressures. The 20 bars limit is scaled by $10^{{\rm log}(g)-5}$ at other surface gravities. The chemistry includes 277 species and out-of-equilibrium chemistry has been performed using a relaxation model\cite{tsailyons2017}. Rainout is assumed to occur for species that are not included in the out-of-equilibrium model. Opacity sources include \ce{H2}-\ce{H2}, \ce{H2}-He, \ce{H2O}, \ce{CO2}, CO, \ce{CH4}, \ce{NH3}, Na, K, Li, Rb, Cs, TiO, VO, FeH, \ce{PH3}, \ce{H2S}, HCN, \ce{C2H2}, \ce{SO2}, Fe, H-, and the Rayleigh scattering opacities for \ce{H2}, He, CO, \ce{N2}, \ce{CH4}, \ce{NH3}, \ce{H2O}, \ce{CO2}, \ce{H2S}, \ce{SO2}. The grids explore the following parameters: effective temperatures between 250 and 1200K; ${\rm log}(g)$ between 2.5 and 5.5 (step 0.5); effective adiabatic index (reddenning) at a value of 1.25. A standard $\chi^2$-minimization procedure was used to find the best fitting model.

\subsection*{petitRADTRANS}
petitRADTRANS, or pRT\cite{molliere_wardenier_2019} (available from \url{https://petitradtrans.readthedocs.io}), is a Python package for the spectral synthesis of exoplanets and allows users to calculate transmission, emission or reflectance spectra. It offers a wide selection of opacities (gas line and continuum, and cloud opacities). Spectra can be calculated at any resolution, up to a wavelength spacing of $\lambda/\Delta \lambda=10^6$. Coupled to a Bayesian inference method such as PyMultiNest\cite{Feroz2008sampling,Buchner2014_BayesianSelection}, 
which pRT provides as a pre-implemented retrieval package, posterior distributions for atmospheric parameters can be derived, given an observation. For WISE~J1828 we assumed a forward model setup as described for the retrievals above, with the priors and forward model details set up as described in the following. The prior on ${\rm log}(g)$ was uniform from 2.5 to 6, while the radius prior ranged from 0.5 to 3 Jupiter radii. In addition, the pressure-temperature profile was parameterized by retrieving temperature values at 10 locations, equidistantly spaced in log-space between $10^{-6}$ and 1000 bar, and then quadratically interpolating the temperature between these nodes in log(pressure). The priors were set up such that the temperature at 1000~bar was uniformly sampled between 100 and 9000~K, and the temperatures at lower-pressure nodes was allowed to be between 0.2-1.0 the temperature of the neighboring deeper atmospheric node. Since kinks in the P-T profiles could be observed in the standard setup, the P-T regularization described above was optionally turned on when deriving the atmospheric model posteriors, fitting both MIRI/MRS and the archival HST WFC3 data. Our constraints on $^{14}$N/$^{15}$N are not affected by the regularization. However, we observed a trend that a regularized P-T leads to a higher atmospheric metal enrichment, higher gravity log(g), smaller radii, and higher effective temperatures, see Extended Data Fig.~2.
For the detection of \ce{$^{15}$NH3} we only used the MRS data. We also turned the regularization off, to allow for maximum model flexibility. This leads to a conservative estimate of the detection significance. The following opacity species were included in the retrievals: \ce{H2O}\cite{Polyansky2018_Exomol}, 
\ce{CH4}\cite{Hargreaves2020_Methane_HITEMP}, 
CO\cite{rothman_2010}, 
\ce{CO2}\cite{exo_co2}, \ce{$^{15}$NH3}\cite{rothman_2013},
\ce{H2S}\cite{exo_h2s}, 
\ce{NH3}\cite{exo_nh3}, and 
\ce{PH3}\cite{exo_ph3}. 
The abundances of said molecules were retrieved using log-uniform priors from $10^{-10}$ to 1 on their mass fractions. For the \ce{$^{15}$NH3} detection we used PyMultiNest, with 2000 live points, \verb|constant_sampling_efficiency| set to False, and a sampling efficiency of 0.3. The detection significance of \ce{$^{15}$NH3} was determined using a standard  method\cite{bennekeseager2013}. We report a detection significance of 4.2~$\sigma$ for \ce{$^{15}$NH3}. For the retrievals constraining the properties of WISE~J1828, which included HST in addition to MRS data, we ran in constant sampling efficiency mode, with the efficiency set to 5 \%. This needed to be done because otherwise retrievals ran for ~$10^8$ models, but did not finish. The partially filled weighted posterior files of the $10^8$-model retrievals were consistent with the results using the constant sampling efficiency mode. With petitRADTRANS we found $^{14}{\rm N}/^{15}{\rm N} = 560_{-115}^{+165}$ for the flexible P-T and $^{14}{\rm N}/^{15}{\rm N} = 642_{-192}^{+365}$ for the regularized P-T model.

\subsection*{ARCiS}
The ARtfull Modeling Code for exoplanet Science (ARCiS) is a forward modelling and retrieval code that can be used to analyse and simulate exoplanet atmosphere spectra. It contains many physical and chemical processes including cloud formation\cite{ormelmin2019} and disequilibrium chemistry\cite{kawashimamin2021}. The free P-T structure used in this work is parameterized by the slope at several pressure points in the atmosphere. We parameterize $d{\rm log}T/d{\rm log}P$ with a prior range between -4/7 to +4/7. The adiabatic gradient expected for a diatomic gas is $d{\rm log}T/d{\rm log}P=2/7$ so this prior range gives a very generous range. We fix the absolute value of the temperature structure at a pressure of $P=0.1$~bar. We retrieve the value of the slope at 12 pressure points equally distributed over the atmosphere in $\log-P$ space. For the detection of \ce{$^{15}$NH3} we follow this procedure, allowing full flexibility and thereby constructing a conservative detection significance. For the final fits deriving the isotopic ratio and the planet parameters presented, we restrict the gradient of the P-T structure to be positive as expected for a non-irradiated atmosphere. Following a standard approach\cite{bennekeseager2013}, we find evidence that \ce{$^{15}$NH3} is present in the atmosphere of WISE~J1828 at 6.3~$\sigma$. We observe the same trend between metal enrichment, ${\rm log}(g)$, radii, and effective temperature as petitRADTRANS when regularizing the P-T structure, see Fig.~2.
However, our derived $^{14}{\rm N}/^{15}{\rm N}$ values are somewhat less stable when turning on regularization. Our regularized values are consistent with the petitRADTRANS (regularized and flexible P-T), namely  $^{14}{\rm N}/^{15}{\rm N} = 591_{-190}^{+432}$. The value derived for the flexible P-T setup is higher ($^{14}{\rm N}/^{15}{\rm N} = 949_{-208}^{+322}$). More information on ARCiS can be found at \url{http://www.exoclouds.com}.

\subsection*{ARCiS+ML}
For the self-consistent retrieval with ARCiS we assume a one-dimensional atmosphere in radiative-convective equilibrium. The retrievals were run on the MIRI MRS data only (i.e., not including the HST data). The atmospheric composition is parameterized using [M/H], C/O and N/O, and we account for disequilibrium chemistry of \ce{CH4}-\ce{H2O}-CO and \ce{NH3}-\ce{N2} due to vertical mixing\cite{kawashimamin2021}, where the vertical eddy diffusion coefficient $K_{zz}$ is another free parameter. For WISE J1828 observed that, although \ce{NH3}-\ce{N2} quenching was active in our models, it quenched from the lowest layer in the atmosphere where NH$_3$ was still the dominating absorber. Likely quenching actually occurs outside our simulated pressure domain, deeper inside the atmosphere. Therefore the ARCiS+ML constraints may be too low for N/O, similar to the constraints from the retrievals, which are insensitive to the spectrally inactive \ce{N2}. The radiative transfer module was benchmarked\cite{chubbmin2022}. Due to the high computational load of the self-consistent models we cannot run nested sampling retrievals, which require millions of model evaluations to converge. Instead, we use a machine learning method based on SNPE (sequential neural posterior estimation\cite{greenberg19a}) that allows us to perform the retrieval using on the order of $10^4$ models. The details of this retrieval method will be presented in an upcoming publication\cite{ardevol23}.

\subsection*{Brewster}
$Brewster$\cite{Burningham2017_Retrieval_CloudyL, Burningham2021_Clouds_2M2244}, is a retrieval code that has mainly been employed in the context of exploring clouds in brown dwarfs. However, here, we utilised a simple cloud-free retrieval recipe. $Brewster$ uses the two stream radiative transfer architecture\cite{Toon1989_RadiativePhotodissociation}.
We use the default 64 layer atmosphere with intervals of 0.1 dex across the pressures range log P(bar) of -4 to 2.3. The temperature is set using P-T parameterisation\cite{Madhusudhan2009_retrieval} 
linked to three atmospheric zones via exponential gradients. We included the molecules \ce{H2O}, \ce{CH4}, \ce{NH3}, \ce{CO}, \ce{PH3} and \ce{H2S} which originate from a compendium\cite{Freedman2008_Opacities,Freedman2014_Opacities}
and updated opacities\cite{Burningham2017_Retrieval_CloudyL}. The abundances of these molecules are modeled assuming vertically constant mixing ratios, which are retrieved as free parameters. The opacities are ingested at a resolution of R=10,000, putting the native model resolution an order of magnitude above the data being fit. Continuum opacity's for H$_2$-H$_2$ and H$_2$-He collisionally induced absorption, Rayleigh scattering due to H$_2$, He and CH$_4$, and continuum opacities due to bound-free and free-free absorption by H- and free-free absorption by H$_2$ are also included. We apply an error inflation "tolerance" framework\cite{lineteske2015}
and used in all subsequent published works using $Brewster$. Only the JWST/MIRI MRS observations were retrieved with $Brewster$, so the HST data was not included. Here we used the EMCEE\cite{ForemanMackey2013_EMCEE} as our sampling algorithm. As we used the standard retrieval recipe for this code, an extensive list of parameter sampling priors has been used\cite{Burningham2021_Clouds_2M2244}.

\section*{Combining model results}
Whenever we report values for the properties of WISE~J1828 in the main body of the text, these have been obtained from combining the posteriors of the retrievals at equal weight, and calculating the corresponding median and 16-84~\%–percentile values, corresponding to the 1~$\sigma$ credible interval in case the resulting distribution is approximately Gaussian. The combined value of the self-consistent codes was obtained by taking the average of their best-fit (ATMO) and median (ARCiS+ML) values, while the uncertainty is obtained from calculating the difference $d$ between these two values, and then calculating $\sqrt{d^2+a^2}$, where $a$ is the ``1~$\sigma$'' uncertainty obtained from the ARCiS+ML posterior.

\section*{Exploring $^{14}{\rm N}/^{15}{\rm N}$ as a formation tracer}

To approximate the impact of volatile ice accretion between the \ce{NH3} and \ce{N2} icelines in a proto-planetary disk, we generated a simplified planet formation model for calculating $^{14}{\rm N}/^{15}{\rm N}$ as a function of the total solid mass (rock and ices) a planet incorporated during formation. For this we used a specific framework\cite{Molliere2022assumptions}, which is available at \url{https://gitlab.com/mauricemolli/formation-inversion}. In short, we used a solar disk composition\cite{Molliere2022assumptions}, their Table~2. This means that the mass ratio between \ce{NH3} and \ce{N2} is 1:7 in the disk (combining gas and ice reservoirs). We then assumed, conservatively, that $^{14}{\rm N}/^{15}{\rm N}$ is reduced by a factor of 2 in \ce{NH3}, when compared to the total $^{14}{\rm N}/^{15}{\rm N}$ value, while the total $^{14}{\rm N}/^{15}{\rm N}$ (summing over all species and phases) is conserved, which we assumed to be 300, and call ISM value in the in Extended Data Fig.~4.
This figure shows the ratio of the planetary and ISM values of $^{14}{\rm N}/^{15}{\rm N}$ as a function of solids accreted by a planet between the \ce{NH3} and \ce{N2} icelines. Since the solids are rich in \ce{NH3}, and \ce{N2} is only in the gas phase, a higher accreted solid mass results in a lower planetary $^{14}{\rm N}/^{15}{\rm N}$. We note that the picture could be more complicated, since the disk gas could be enriched in $^{15}{\rm N}$-poor \ce{N2} gas that evaporated off pebbles that drift in from outside the \ce{N2} iceline\cite{schneiderbitsch2021}.

\end{methods}



\begin{addendum}

\item[Data availability]

{The JWST MIRI data presented in this paper are part of the JWST MIRI GTO program (Program identifier (PID) 1189; P.I.\  T. Roellig). The JWST data will be publicly available in the Barbara A.\ Mikulski Archive for Space Telescopes (MAST; \url{https://archive.stsci.edu/}) after July 28th, 2023, and can be found either using the program identifier or using the Data Object Identifier (DOI): \url{https://doi.org/10.17909/as3s-x893}. 
The HST WFC3 spectrum is available from: \url{https://cdsarc.cds.unistra.fr/viz-bin/cat/J/ApJ/920/20#/article}. 
}

\item[Code availability]
The codes used in this publication to extract, reduce, and analyse the data are as follows; The data reduction pipeline \texttt{jwst} can be found at \url{https://jwst-pipeline.readthedocs.io/en/latest/}. The atmospheric model codes used to fit the data can be found at \url{https://www.exoclouds.com/} for the 
\texttt{ARCiS}-code\,\cite{min_2020} and at
\url{https://petitradtrans.readthedocs.io/en/latest/} for the \texttt{petitRADTRANS}-code\,\cite{molliere_wardenier_2019}. The simplified planet formation model\cite{Molliere2022assumptions} used to study $^{14}$N/$^{15}$N  as a function of accreted ice mass can be found at \url{https://gitlab.com/mauricemolli/formation-inversion}. The detailed setups of the open source tools for the analyses presented here are described in the methods section of this paper, and can be made available to interested parties upon request.

\item[Inclusion \& Ethics] All authors have committed to upholding the principles of research ethics \& inclusion as advocated by the Nature Portfolio journals.  

\item [Acknowledgments] This work is based [in part] on observations made with the NASA/ESA/CSA James Webb Space Telescope. The data were obtained from the Mikulski Archive for Space Telescopes at the Space Telescope Science Institute, which is operated by the Association of Universities for Research in Astronomy, Inc., under NASA contract NAS 5-03127 for JWST. These observations are associated with program 1189. MIRI draws on the scientific and technical expertise of the following organisations: Ames Research Center, USA; Airbus Defence and Space, UK; CEA-Irfu, Saclay, France; Centre Spatial de Li\`ege, Belgium; Consejo Superior de Investigaciones Cient\'{\i}ficas, Spain; Carl Zeiss Optronics, Germany; Chalmers University of Technology, Sweden; Danish Space Research Institute, Denmark; Dublin Institute for Advanced Studies, Ireland; European Space Agency, Netherlands; ETCA, Belgium; ETH Z\"urich, Switzerland; Goddard Space Flight Center, USA; Institut d'Astrophysique Spatiale, France; Instituto Nacional de T\'ecnica Aeroespacial, Spain; Institute for Astronomy, Edinburgh, UK; Jet Propulsion Laboratory, USA; Laboratoire d'Astrophysique de Marseille (LAM), France; Leiden University, Netherlands; Lockheed Advanced Technology Center (USA); NOVA Opt-IR group at Dwingeloo, Netherlands; Northrop Grumman, USA; Max-Planck Institut f\"ur Astronomie (MPIA), Heidelberg, Germany; Laboratoire d'Etudes Spatiales et d'Instrumentation en Astrophysique (LESIA), France; Paul Scherrer Institut, Switzerland; Raytheon Vision Systems, USA; RUAG Aerospace, Switzerland; Rutherford Appleton Laboratory (RAL Space), UK; Space Telescope Science Institute, USA; Toegepast Natuurwetenschappelijk Onderzoek (TNO-TPD), Netherlands; UK Astronomy Technology Centre, UK; University College London, UK; University of Amsterdam, Netherlands; University of Arizona, USA; University of Bern, Switzerland; University of Cardiff, UK; University of Cologne, Germany; University of Ghent; University of Groningen, Netherlands; University of Leicester, UK; KU Leuven, Belgium; University of Stockholm, Sweden; Utah, State University, USA. The following National and International Funding Agencies
funded and supported the MIRI development:
NASA; ESA; Belgian Science Policy Office (BELSPO);
Centre Nationale d'Etudes Spatiales (CNES); Danish
National Space Centre; Deutsches Zentrum fur Luftund
Raumfahrt (DLR); Enterprise Ireland; Ministerio
de Econom\'{\i}a y Competividad; Netherlands Research
School for Astronomy (NOVA); Netherlands Organisation
for Scientific Research (NWO); Science and Technology
Facilities Council; Swiss Space Office; Swedish
National Space Agency; and UK Space Agency. 
D.B.\ and M.M.C. are supported by Spanish MCIN/AEI/10.13039/501100011033 grant PID2019-107061GB-C61 and and No. MDM-2017-0737. 
C.C.\, A.B.\, P.-O.L.\, R.G.\, A.C.\ acknowledge funding support from CNES. 
P.P.\ thanks the Swiss National Science Foundation (SNSF) for financial support under grant number 200020\_200399.
N.W.\ acknowledges funding from NSF Award 1909776 and NASA XRP Award 80NSSC22K0142.
O.A.\, I.A.\, B.V.\ and P.R.\ thank the European Space Agency (ESA) and the Belgian Federal Science Policy Office (BELSPO) for their support in the framework of the PRODEX Programme.
L.D.\ acknowledges funding from the KU Leuven Interdisciplinary Grant (IDN/19/028), the European Union H2020-MSCA-ITN-2019 under Grant no. 860470 (CHAMELEON) and the FWO research grant G086217N. 
I.K.\ acknowledges support from grant TOP-1 614.001.751 from the Dutch Research Council (NWO). 
O.K.\ acknowledges support from the Federal Ministery of Economy (BMWi) through the German Space Agency (DLR).
J.P.P.\ acknowledges financial support from the UK Science and Technology Facilities Council, and the UK Space Agency.
G. O \ acknowledges support from the Swedish National Space Board and the Knut and Alice Wallenberg Foundation. 
P.T.\ acknowledges support by the European Research Council under Grant Agreement ATMO 757858. 
F.A.M.\ has received funding from the European Union's Horizon 2020 research and innovation programme under the Marie Sk\l{}odowska-Curie grant agreement no.\ 860470. 
L.C.\ acknowledges support by grant PIB2021-127718NB-100 from the Spanish Ministry of Science and Innovation/State Agency of Research MCIN/AEI/10.13039/501100011033. 
E.vD.\ acknowledges support from A-ERC grant 101019751 MOLDISK. 
T.P.R.\ acknowledges support from the ERC 743029 EASY. 
G.\"O.\ acknowledges support from SNSA.
Th.H. acknowledge support from the European Research Council under the European Union’s Horizon 2020 research and innovation program under grant agreement No. 832428-Origins.
We thank the MIRI instrument team and the many others who contributed to the success of JWST.

\item[Author Contributions] 
All authors played a significant role in one or more of the following: designing and building MIRI, development of the original proposal, management of the project, definition of the target list and observation plan, analysis of the data, theoretical modelling and preparation of this paper. Some specific contributions are listed as follows. 
P.-O.L.\ is PI of the JWST MIRI GTO European consortium program dedicated to JWST observations of exoplanet atmospheres. 
D.B., P.M., and P.P. provided overall program leadership and management of the WISE~J1828 working group.
P.-O.L., T.H., R.W., (co-lead of the JWST MIRI GTO European consortium), D.B. and M.Mu. made significant contributions to the design of the observational program and contributed to the setting of the observing parameters. 
P.P., I.A., and M.S. \ reduced the data. 
P.T. generated theoretical model grids for comparison with the data. 
P.M., M.Mi., and N.W.\ fitted the generated spectrum with retrieval models, and M.V. also contributed.
F.A.M. \ applied the radiative-convective equilibrium retrieval to the spectrum.
D.B, P.M., and P.P.\ led the writing of the manuscript. 
P.-O.L. and L.D. \ made significant contributions to the writing of this paper. Additional contribution was provided by M.M.C. and L.D.
P.B., A.B., J.B., C.C., A.C., N.C., R.G., A.G., A.M.G., S.K., F.L., D.R., P.R., S.S., and I.W. contributed to instrument construction, the program design and/or the data analysis. 
P.M., P.P. and D.B.\  generated figures for this paper.
G.W.\ is PI of the JWST MIRI instrument,
P.-O.L., T.H., M.G, B.V., L.C., E.vD., T.G., T.R., G.Os.\ are co-PIs, and
L.D., R.W., O.A., I.K., O.K., J.P., G.Ol.\, and D.B. are co-Is of the JWST MIRI instrument. 

 \item[Competing Interests] The authors declare that they have no competing financial interests.
 
 \item[Correspondence] Correspondence and requests for materials should be addressed to barrado@cab.inta-csic.es.

Reprints and permissions information is available at www.nature.com/reprints
\end{addendum}
\afterpage{\clearpage}
\newpage

\setcounter{figure}{0} 
\renewcommand{\figurename}{Extended Data -- Figure}


\setcounter{table}{0} 
\renewcommand{\tablename}{Extended Data -- Table}

\section*{Extended Data}

\begin{figure*}[!ht]
\centering
	\includegraphics[width=\textwidth]{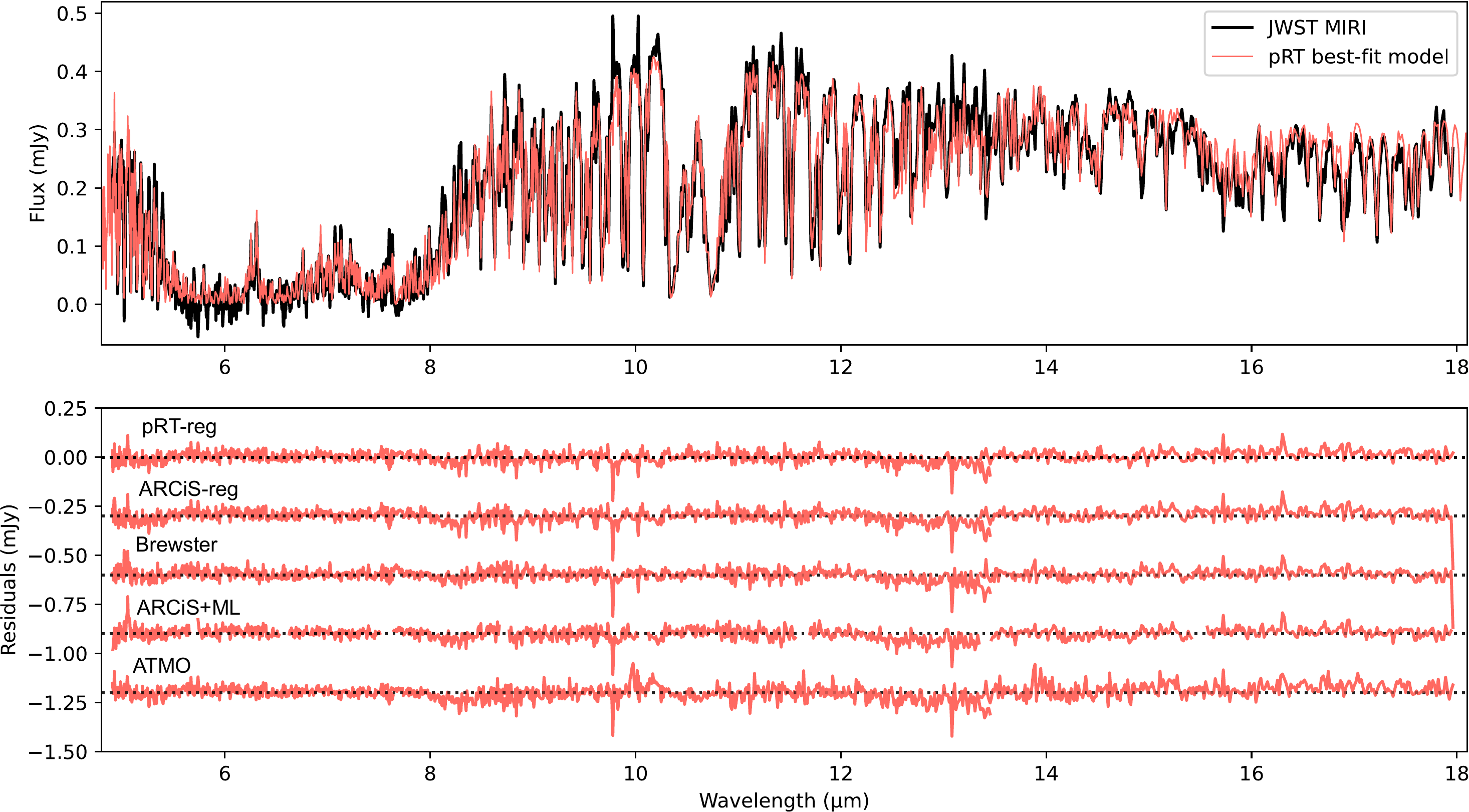}
	\caption{\textbf{The spectrum of WISE~J1828 (black solid lines) and best-fit model.} We show the MIRI/MRS spectrum of WISE~J1828 (black solid lines) and best-fit model of the regularized-P-T retrieval of petitRADTRANS (red line). Residuals (models $-$ observed spectrum) are displayed at the bottom panel. pRT-reg and ARCiS-reg stand for the regularized P-T retrieval of petitRADTRANS and ARCiS, respectively.}
		\label{fig:spectrum_retrieval}
\end{figure*}

\begin{figure*}[!ht]
\centering
	\includegraphics[width=\textwidth]{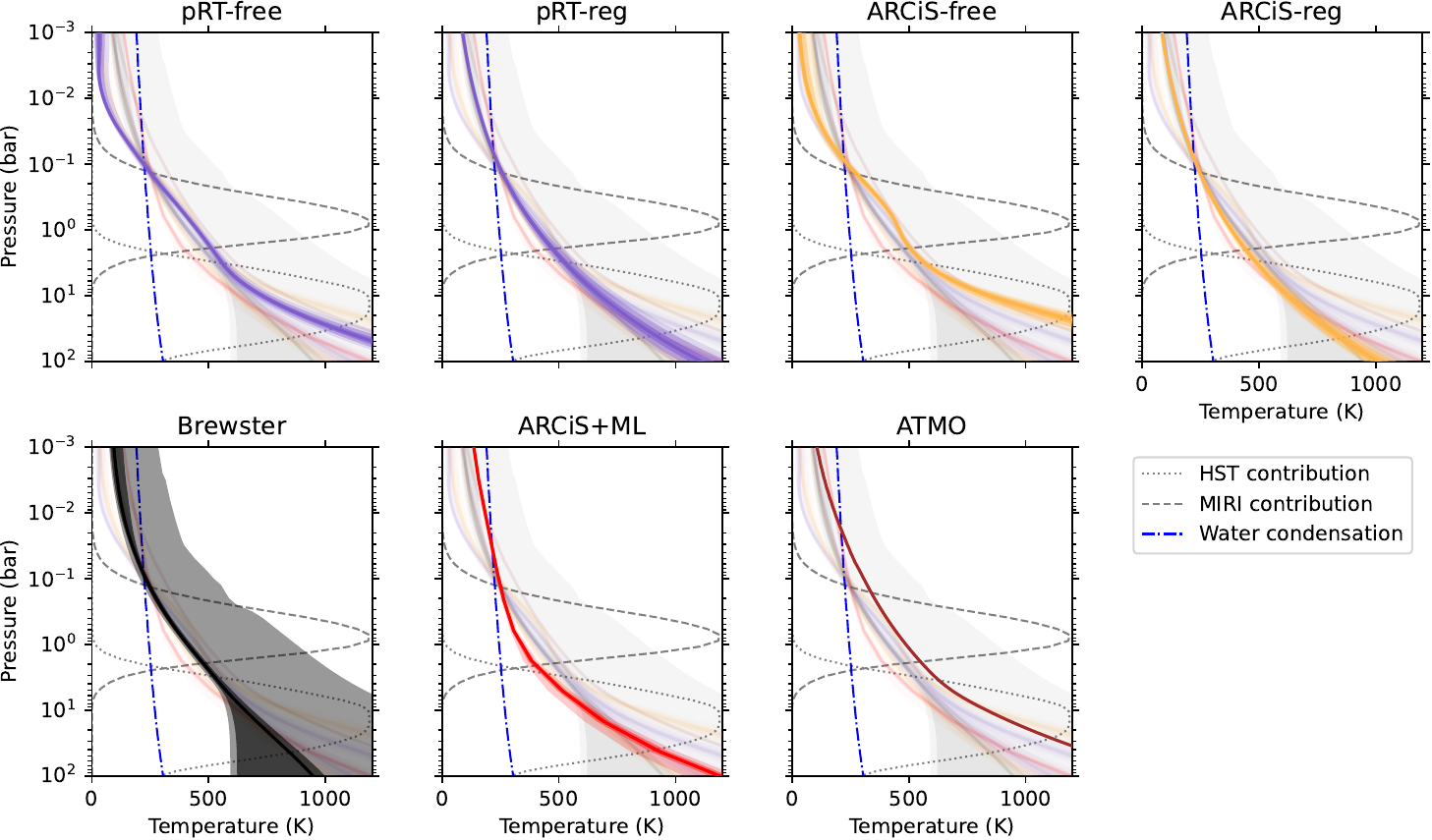}
	\caption{\textbf{Model inferences on the various P-T profiles derived for WISE~J1828.} The individual panels always highlight the constraint from one given model, while the results of the other models are shown in the background. pRT-reg and ARCiS-reg stand for the regularized P-T retrievals, while pRT-free and ARCiS-free stand for the unregularized P-T retrievals of petitRADTRANS and ARCiS, respectively. The contribution functions of the HST and MIRI observations, constrained from the best-fit pRT-reg model, are shown as dotted and dashed lines, respectively. The condensation curve for water (at solar metallicity) is shown as a blue dashdotted curve., indicating that, while neglected in our models, water clouds could impact the spectrum in a modest away. }
		\label{fig:pt_profiles}
\end{figure*}

\begin{figure*}[!ht]
\centering
	\includegraphics[width=\textwidth]{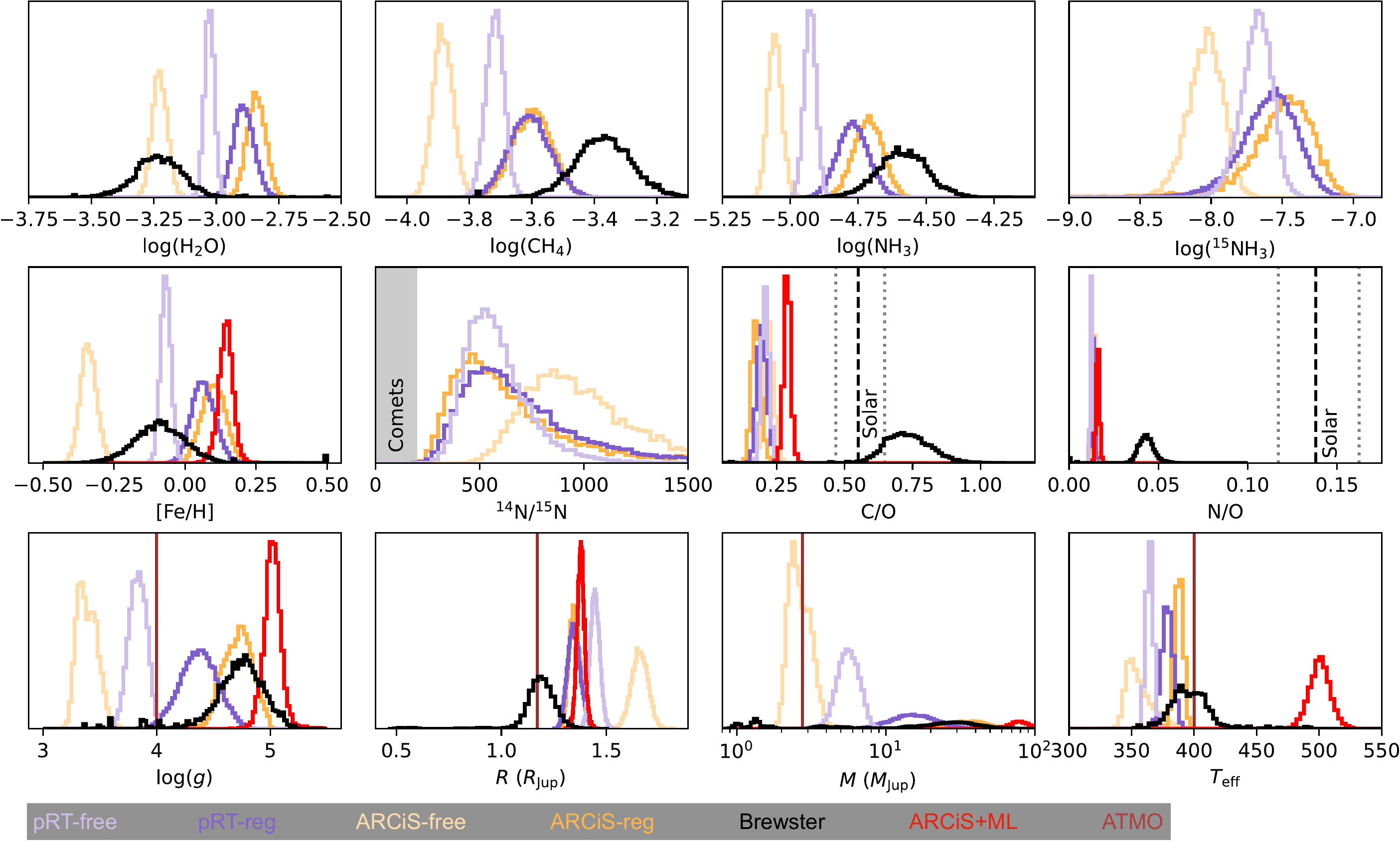}
	\caption{\textbf{One-dimensional projection of the posterior distributions of the WISE~J1828 retrievals.} Values correspond to key atmospheric quantities shown in Extended Data Table~1.
          pRT-reg and ARCiS-reg stand for the regularized P-T retrievals, while pRT-free and ARCiS-free stand for the unregularized P-T retrievals of petitRADTRANS and ARCiS, respectively.}
		\label{fig:posteriors}
\end{figure*}

\begin{figure*}[!ht]
\centering
	\includegraphics[width=\textwidth]{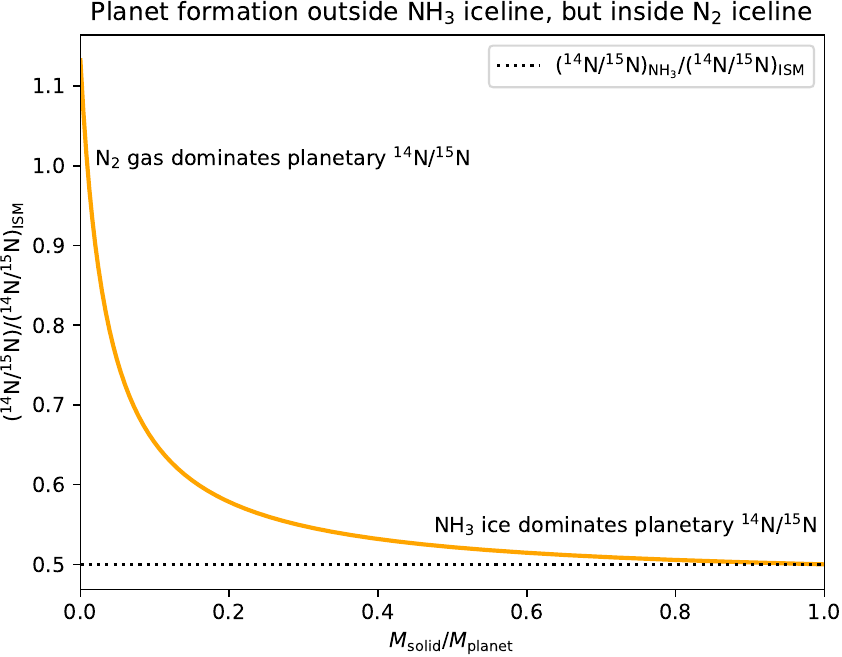}
	\caption{\textbf{Evolution of the planetary $^{14}$N/$^{15}$N as a function of the mass accreted in solids (rock and ice).} This computation assumes a planet that forms outside the \ce{NH3} ice line but inside the \ce{N2} ice line. The black dotted line denotes the value expected for pure \ce{NH3} ice.}
		\label{fig:toy_model}
\end{figure*}

\thispagestyle{empty}

\begin{table*}[!htp]
	\caption{\textbf{Physical constraints on WISE~J1828}}
                {``free'' in the code name means that the P-T structure was not regularized, while this was done for the ``reg'' cases. $R_{\rm bin}$ model radii have been multiplied by $1/\sqrt{2}$, assuming WISE J1828 is an equal-property binary. $R_{\rm bin}$ was used for calculating $M_{\rm bin}$ from the inferred gravity. The units of $K_{zz}$ are cm$^2$ s$^{-1}$. ``chem'' means that absorber abundances have been determined from a chemical model. ``solar'' means that the parameter was not varied, and that a solar composition was assumed instead.\\
                }
	\label{tab:retrieval_result}
	\begin{center}
	\begin{tabular}{@{\extracolsep\fill}lcccccccc}
		\toprule%
		& \multicolumn{5}{@{}c@{}}{Retrieval codes} & & \multicolumn{2}{@{}c@{}}{Self-consistent codes} \\
        \cmidrule{2-6} \cmidrule{8-9}%
		Quantity & pRT-free & pRT-reg & ARCiS-free & ARCiS-reg & Brewster & & ARCiS+ML & ATMO  \\
		\midrule
		$T_{\rm eff}$ (K) & $364_{-3}^{+3}$ & $379_{-4}^{+4}$ & $354_{-7}^{+13}$ & $388_{-4}^{+3}$ & $397_{-14}^{+16}$ & & $501_{-8}^{+8}$ & 400 \\
        ${\rm log}(g)$ & $3.83_{-0.08}^{+0.08}$ & $4.37_{-0.17}^{+0.16}$ & $3.38_{-0.08}^{+0.1}$ & $4.72_{-0.14}^{+0.12}$ & $4.74_{-0.21}^{+0.19}$ & & $5.02_{-0.07}^{+0.07}$ & 4.0 \\
        $R$ ($\rm R_{Jup}$) & $1.45_{-0.02}^{+0.02}$ & $1.35_{-0.03}^{+0.03}$ & $1.65_{-0.04}^{+0.04}$ & $1.35_{-0.03}^{+0.03}$ & $1.19_{-0.06}^{+0.06}$ & & $1.38_{-0.02}^{+0.02}$ & 1.17 \\
        $M$ ($\rm M_{Jup}$) & $5.76_{-0.89}^{+1.04}$ & $17.27_{-5.01}^{+6.86}$ & $2.69_{-0.40}^{+0.58}$ & $38.57_{-9.66}^{+10.92}$ & $31.50_{-11.60}^{+14.30}$ & & $79.99_{-10.87}^{+12.79}$ & 2.76 \\
        $R_{\rm bin}$ ($\rm R_{Jup}$) & $1.02_{-0.02}^{+0.02}$ & $0.95_{-0.02}^{+0.02}$ & $1.17_{-0.03}^{+0.03}$ & $0.95_{-0.02}^{+0.02}$ & $0.84_{-0.04}^{+0.04}$ & & $0.98_{-0.01}^{+0.01}$ & 0.83  \\
        $M_{\rm bin}$ ($\rm M_{Jup}$) & $2.88_{-0.45}^{+0.52}$ & $8.63_{-2.51}^{+3.43}$ & $1.34_{-0.2}^{+0.29}$ & $19.29_{-4.83}^{+5.46}$ & $15.75_{-5.8}^{+7.15}$ & & $39.99_{-5.43}^{+6.4}$ & 1.38\\
        \midrule
        $\rm [M/H]$ & $-0.07_{-0.02}^{+0.02}$ & $0.06_{-0.04}^{+0.04}$ & $-0.34_{-0.03}^{+0.03}$ & $0.10_{-0.04}^{+0.04}$ & $-0.08_{-0.09}^{+0.10}$ & & $0.14_{-0.03}^{+0.02}$ & solar \\
        C/O & $0.21_{-0.01}^{+0.01}$ & $0.20_{-0.02}^{+0.02}$ & $0.22_{-0.02}^{+0.02}$ & $0.17_{-0.02}^{+0.02}$ & $0.73_{-0.10}^{+0.08}$  & & $0.29_{-0.01}^{+0.01}$ & solar \\
        (N/O) \ / & $0.091_{-0.004}^{+0.004}$ & $0.096_{-0.006}^{+0.006}$ & $0.107_{-0.006}^{+0.006}$ & $0.097_{-0.006}^{+0.006}$ & $0.314_{-0.032}^{+0.036}$
  & & $0.117_{-0.007}^{+0.006}$ & solar \\
              \ \ \ (N/O)$_\odot$\\
        $^{14}{\rm N}/^{15}{\rm N}$ & $560_{-115}^{+165}$ & $642_{-192}^{+365}$ & $949_{-208}^{+322}$ & $591_{-190}^{+432}$ & -- & & -- & -- \\
        \midrule
        log(\ce{H2O}) & $-3.03_{-0.02}^{+0.02}$ & $-2.89_{-0.04}^{+0.04}$ &  $-3.23_{-0.03}^{+0.03}$ & $-2.84_{-0.04}^{+0.04}$ & $-3.23_{-0.10}^{+0.10}$ & & chem & chem\\
        log(\ce{CH4}) & $-3.72_{-0.03}^{+0.03}$ & $-3.61_{-0.06}^{+0.07}$ & $-3.88_{-0.03}^{+0.03}$ & $-3.60_{-0.06}^{+0.06}$ & $-3.36_{-0.09}^{+0.10}$ & & chem & chem \\
        log(\ce{NH3}) & $-4.93_{-0.02}^{+0.02}$ & $-4.77_{-0.06}^{+0.06}$ & $-5.06_{-0.02}^{+0.03}$ & $-4.71_{-0.05}^{+0.05}$ & $-4.58_{-0.10}^{+0.10}$ & & chem & chem \\
        log(\ce{^{15}NH3}) & $-7.67_{-0.11}^{+0.10}$ & $-7.58_{-0.21}^{+0.17}$ & $-8.03_{-0.13}^{+0.11}$ & $-7.49_{-0.25}^{+0.18}$ & -- & & -- & -- \\
        \midrule
        log($K_{\rm zz}$) & -- & -- & -- & -- & -- & & $9.01_{-0.19}^{+0.19}$ & 7 \\
		\bottomrule
	\end{tabular}\\
\end{center}
\end{table*}

\newpage

\thispagestyle{empty}

\begin{table*}[!htp]
	\caption{\textbf{Physical constraints on WISE~J1828, combining different codes}}
                {$R_{\rm bin}$ model radii have been multiplied by $1/\sqrt{2}$, assuming WISE J1828 is an equal-property binary. $R_{\rm bin}$ was used for calculating $M_{\rm bin}$ from the inferred gravity. The units of $K_{zz}$ are cm$^2$ s$^{-1}$. ``chem'' means that absorber abundances have been determined from a chemical model. ``no comb'' means that the respective parameter was only varied in one of the two codes, please see Extended Data Table~1 
                  for the inferred values. $^*$ for the inferred average mass denotes that the distance between the two codes' solutions was larger than the average value, so no uncertainties, derived as explained in the method section, are given here.\\
                }
	\label{tab:retrieval_result_combined}
	\begin{center}
	\begin{tabular}{@{\extracolsep\fill}lcc}
		\toprule%
		Quantity & Retrievals & Self-consistent codes \\
		\midrule
		$T_{\rm eff}$ (K) & $378_{-18}^{+13}$ & $450\pm 101$
\\
        ${\rm log}(g)$ & $4.34_{-0.88}^{+0.42}$ & $4.5\pm 1.0$  \\
        $R$ ($\rm R_{Jup}$) & $1.37_{-0.13}^{+0.26}$ & $1.27\pm 0.21$
 \\
        $M$ ($\rm M_{Jup}$) & $15.8_{-12.64}^{+22.82}$ &  $41^*$\\
        $R_{\rm bin}$ ($\rm R_{Jup}$) & $0.97_{-0.09}^{+0.18}$ & $0.90\pm 0.15$ \\
        $M_{\rm bin}$ ($\rm M_{Jup}$) & $7.91_{-6.34}^{+11.33}$ & $21^*$ \\
        \midrule
        $\rm [M/H]$ & $-0.05_{-0.27}^{+0.15}$ & no comb  \\
        C/O & $0.21_{-0.03}^{+0.45}$ & no comb \\
        (N/O) \ / & $0.10_{-0.01}^{+0.19}$ & no comb \\
              \ \ \ (N/O)$_\odot$\\
        $^{14}{\rm N}/^{15}{\rm N}$ & $673_{-212}^{+393}$ & -- \\
        \midrule
        log(\ce{H2O}) & $-3.03_{-0.21}^{+0.18}$ & chem \\
        log(\ce{CH4}) & $-3.65_{-0.21}^{+0.21}$ & chem \\
        log(\ce{NH3}) & $-4.79_{-0.25}^{+0.15}$ & chem \\ log(\ce{^{15}NH3}) & $-7.68_{-0.34}^{+0.24}$ & -- \\
        \midrule
        log($K_{\rm zz}$) & -- & $8\pm2$ \\
		\bottomrule
	\end{tabular}\\
\end{center}
\end{table*}

\newpage

\afterpage{\clearpage}
\newpage


\begin{thebibliography}{10}
\expandafter\ifx\csname url\endcsname\relax
  \def\url#1{\texttt{#1}}\fi
\expandafter\ifx\csname urlprefix\endcsname\relax\def\urlprefix{URL }\fi
\providecommand{\bibinfo}[2]{#2}
\providecommand{\eprint}[2][]{\url{#2}}

\bibitem{burrows1997nongray}
\bibinfo{author}{Burrows, A.} \emph{et~al.}
\newblock \bibinfo{title}{A nongray theory of extrasolar giant planets and
  brown dwarfs}.
\newblock \emph{\bibinfo{journal}{Astrophys. J.}}
  \textbf{\bibinfo{volume}{491}}, \bibinfo{pages}{856--875}
  (\bibinfo{year}{1997}).

\bibitem{faherty2018spectral}
\bibinfo{author}{Faherty, J.~K.}
\newblock \bibinfo{title}{Spectral properties of brown dwarfs and unbound
  planetary mass objects}.
\newblock In \bibinfo{editor}{Deeg, H.} \& \bibinfo{editor}{Belmonte, J.}
  (eds.) \emph{\bibinfo{booktitle}{Handbook of Exoplanets}}
  (\bibinfo{publisher}{Springer}, \bibinfo{address}{Cham},
  \bibinfo{year}{2018}).

\bibitem{Madhusudhan2014migration}
\bibinfo{author}{{Madhusudhan}, N.}, \bibinfo{author}{{Amin}, M.~A.} \&
  \bibinfo{author}{{Kennedy}, G.~M.}
\newblock \bibinfo{title}{{Toward Chemical Constraints on Hot Jupiter
  Migration}}.
\newblock \emph{\bibinfo{journal}{Astrophys. J. Letters}}
  \textbf{\bibinfo{volume}{794}}, \bibinfo{pages}{L12} (\bibinfo{year}{2014}).

\bibitem{Molliere2022assumptions}
\bibinfo{author}{{Molli{\`e}re}, P.} \emph{et~al.}
\newblock \bibinfo{title}{{Interpreting the Atmospheric Composition of
  Exoplanets: Sensitivity to Planet Formation Assumptions}}.
\newblock \emph{\bibinfo{journal}{Astrophys. J.}}
  \textbf{\bibinfo{volume}{934}}, \bibinfo{pages}{74} (\bibinfo{year}{2022}).

\bibitem{Feuchtgruber2013_D_H_UranusNeptune}
\bibinfo{author}{{Feuchtgruber}, H.} \emph{et~al.}
\newblock \bibinfo{title}{{The D/H ratio in the atmospheres of Uranus and
  Neptune from Herschel-PACS observations}}.
\newblock \emph{\bibinfo{journal}{Astron. Astrophys.}}
  \textbf{\bibinfo{volume}{551}}, \bibinfo{pages}{A126} (\bibinfo{year}{2013}).

\bibitem{Alibert2018NatAs...2..873A}
\bibinfo{author}{{Alibert}, Y.} \emph{et~al.}
\newblock \bibinfo{title}{{The formation of Jupiter by hybrid
  pebble-planetesimal accretion}}.
\newblock \emph{\bibinfo{journal}{Nature Astronomy}}
  \textbf{\bibinfo{volume}{2}}, \bibinfo{pages}{873--877}
  (\bibinfo{year}{2018}).

\bibitem{nomura2022isotopic}
\bibinfo{author}{Nomura, H.}, \bibinfo{author}{Furuya, K.},
  \bibinfo{author}{Cordiner, M.~A.} \emph{et~al.}
\newblock \bibinfo{title}{The isotopic links from planet formation regions to
  the solar system}.
\newblock In \emph{\bibinfo{booktitle}{Protostars and Planets VII}}
  (\bibinfo{year}{2022}).
\newblock \bibinfo{note}{Retrieved from astro-ph/10863v1}.

\bibitem{zhang2021a}
\bibinfo{author}{Zhang, Y.}, \bibinfo{author}{Snellen, I. A.~G.},
  \bibinfo{author}{Bohn, A.~J.} \emph{et~al.}
\newblock \bibinfo{title}{The 13co-rich atmosphere of a young accreting
  super-jupiter}.
\newblock \emph{\bibinfo{journal}{Nature}} \textbf{\bibinfo{volume}{595}},
  \bibinfo{pages}{370--372} (\bibinfo{year}{2021}).

\bibitem{Line2021C_O_HotAtmos}
\bibinfo{author}{{Line}, M.~R.} \emph{et~al.}
\newblock \bibinfo{title}{{A solar C/O and sub-solar metallicity in a hot
  Jupiter atmosphere}}.
\newblock \emph{\bibinfo{journal}{Nature}} \textbf{\bibinfo{volume}{598}},
  \bibinfo{pages}{580--584} (\bibinfo{year}{2021}).

\bibitem{adande2012millimeter}
\bibinfo{author}{Adande, G.~R.} \& \bibinfo{author}{Ziurys, L.~M.}
\newblock \bibinfo{title}{Millimeter-wave observations of cn and hnc and their
  15n isotopologues: a new evaluation of the 14n/15n ratio across the galaxy}.
\newblock \emph{\bibinfo{journal}{Astrophys. J.}}
  \textbf{\bibinfo{volume}{744}}, \bibinfo{pages}{194} (\bibinfo{year}{2012}).

\bibitem{cushing2011discovery}
\bibinfo{author}{Cushing, M.~C.} \emph{et~al.}
\newblock \bibinfo{title}{The discovery of y dwarfs using data from the
  wide-field infrared survey explorer (wise)}.
\newblock \emph{\bibinfo{journal}{Astrophys. J.}}
  \textbf{\bibinfo{volume}{743}}, \bibinfo{pages}{50} (\bibinfo{year}{2011}).

\bibitem{Morley2018_Lspectrum_Ydwarf}
\bibinfo{author}{{Morley}, C.~V.} \emph{et~al.}
\newblock \bibinfo{title}{{An L Band Spectrum of the Coldest Brown Dwarf}}.
\newblock \emph{\bibinfo{journal}{Astrophys. J.}}
  \textbf{\bibinfo{volume}{858}}, \bibinfo{pages}{97} (\bibinfo{year}{2018}).

\bibitem{Skemer2016_Spectra_Ydwarfs}
\bibinfo{author}{{Skemer}, A.~J.} \emph{et~al.}
\newblock \bibinfo{title}{{The First Spectrum of the Coldest Brown Dwarf}}.
\newblock \emph{\bibinfo{journal}{Astrophys. J. Letters}}
  \textbf{\bibinfo{volume}{826}}, \bibinfo{pages}{L17} (\bibinfo{year}{2016}).

\bibitem{Cushing2021_J1828}
\bibinfo{author}{{Cushing}, M.~C.} \emph{et~al.}
\newblock \bibinfo{title}{{An Improved Near-infrared Spectrum of the Archetype
  Y Dwarf WISEP J182831.08+265037.8}}.
\newblock \emph{\bibinfo{journal}{Astrophys. J.}}
  \textbf{\bibinfo{volume}{920}}, \bibinfo{pages}{20} (\bibinfo{year}{2021}).

\bibitem{beilercushing2023}
\bibinfo{author}{{Beiler}, S.} \emph{et~al.}
\newblock \bibinfo{title}{{The First JWST Spectral Energy Distribution of a Y
  dwarf}}.
\newblock \emph{\bibinfo{journal}{arXiv e-prints}}
  \bibinfo{pages}{arXiv:2306.11807} (\bibinfo{year}{2023}).

\bibitem{Argyriou2023_MRS}
\bibinfo{author}{{Argyriou}, I.} \emph{et~al.}
\newblock \bibinfo{title}{{JWST MIRI flight performance: The Medium-Resolution
  Spectrometer}}.
\newblock \emph{\bibinfo{journal}{arXiv e-prints}}
  \bibinfo{pages}{arXiv:2303.13469} (\bibinfo{year}{2023}).

\bibitem{Burningham2017_Retrieval}
\bibinfo{author}{{Burningham}, B.} \emph{et~al.}
\newblock \bibinfo{title}{{Retrieval of atmospheric properties of cloudy L
  dwarfs}}.
\newblock \emph{\bibinfo{journal}{Mon. Not. R. Astron. Soc.}}
  \textbf{\bibinfo{volume}{470}}, \bibinfo{pages}{1177--1197}
  (\bibinfo{year}{2017}).

\bibitem{molliere_wardenier_2019}
\bibinfo{author}{{Molli{\`e}re}, P.} \emph{et~al.}
\newblock \bibinfo{title}{{petitRADTRANS. A Python radiative transfer package
  for exoplanet characterization and retrieval}}.
\newblock \emph{\bibinfo{journal}{Astron. Astrophys.}}
  \textbf{\bibinfo{volume}{627}}, \bibinfo{pages}{A67} (\bibinfo{year}{2019}).

\bibitem{min_2020}
\bibinfo{author}{{Min}, M.}, \bibinfo{author}{{Ormel}, C.~W.},
  \bibinfo{author}{{Chubb}, K.}, \bibinfo{author}{{Helling}, C.} \&
  \bibinfo{author}{{Kawashima}, Y.}
\newblock \bibinfo{title}{{The ARCiS framework for exoplanet atmospheres.
  Modelling philosophy and retrieval}}.
\newblock \emph{\bibinfo{journal}{Astron. Astrophys.}}
  \textbf{\bibinfo{volume}{642}}, \bibinfo{pages}{A28} (\bibinfo{year}{2020}).

\bibitem{Tremblin2015Fingering}
\bibinfo{author}{{Tremblin}, P.} \emph{et~al.}
\newblock \bibinfo{title}{{Fingering Convection and Cloudless Models for Cool
  Brown Dwarf Atmospheres}}.
\newblock \emph{\bibinfo{journal}{Astrophys. J. Letters}}
  \textbf{\bibinfo{volume}{804}}, \bibinfo{pages}{L17} (\bibinfo{year}{2015}).

\bibitem{chubbmin2022}
\bibinfo{author}{{Chubb}, K.~L.} \& \bibinfo{author}{{Min}, M.}
\newblock \bibinfo{title}{{Exoplanet Atmosphere Retrievals in 3D Using Phase
  Curve Data with ARCiS: Application to WASP-43b}}.
\newblock \emph{\bibinfo{journal}{Astron. Astrophys.}}
  \textbf{\bibinfo{volume}{665}}, \bibinfo{pages}{A2} (\bibinfo{year}{2022}).

\bibitem{ardevol23}
\bibinfo{author}{Ard\'evol~Mart\'inez, F.}, \bibinfo{author}{Min, M.},
  \bibinfo{author}{Huppenkothen, D.}, \bibinfo{author}{Palmer, P.~I.} \&
  \bibinfo{author}{Kamp, I.}
\newblock \bibinfo{title}{Floppity: A new machine learning framework for
  exoplanet atmospheric retrievals.}
\newblock \emph{\bibinfo{journal}{In preparation}} .

\bibitem{defurio2023jwst}
\bibinfo{author}{{De Furio}, M.} \emph{et~al.}
\newblock \bibinfo{title}{{JWST Observations of the Enigmatic Y-Dwarf WISE
  1828+2650. I. Limits to a Binary Companion}}.
\newblock \emph{\bibinfo{journal}{Astrophys. J.}}
  \textbf{\bibinfo{volume}{948}}, \bibinfo{pages}{92} (\bibinfo{year}{2023}).

\bibitem{Phillips2020_Models_TYdwarfs}
\bibinfo{author}{{Phillips}, M.~W.} \emph{et~al.}
\newblock \bibinfo{title}{{A new set of atmosphere and evolution models for
  cool T-Y brown dwarfs and giant exoplanets}}.
\newblock \emph{\bibinfo{journal}{Astron. Astrophys.}}
  \textbf{\bibinfo{volume}{637}}, \bibinfo{pages}{A38} (\bibinfo{year}{2020}).

\bibitem{Asplund2009_Sun_COmposition}
\bibinfo{author}{{Asplund}, M.}, \bibinfo{author}{{Grevesse}, N.},
  \bibinfo{author}{{Sauval}, A.~J.} \& \bibinfo{author}{{Scott}, P.}
\newblock \bibinfo{title}{{The Chemical Composition of the Sun}}.
\newblock \emph{\bibinfo{journal}{Annual Review of Astron. Astrophys.}}
  \textbf{\bibinfo{volume}{47}}, \bibinfo{pages}{481--522}
  (\bibinfo{year}{2009}).

\bibitem{Zahnle2014_CH4_CO_NH3_BD}
\bibinfo{author}{{Zahnle}, K.~J.} \& \bibinfo{author}{{Marley}, M.~S.}
\newblock \bibinfo{title}{{Methane, Carbon Monoxide, and Ammonia in Brown
  Dwarfs and Self-Luminous Giant Planets}}.
\newblock \emph{\bibinfo{journal}{Astrophys. J.}}
  \textbf{\bibinfo{volume}{797}}, \bibinfo{pages}{41} (\bibinfo{year}{2014}).

\bibitem{milesskemer2020}
\bibinfo{author}{{Miles}, B.~E.} \emph{et~al.}
\newblock \bibinfo{title}{{Observations of Disequilibrium CO Chemistry in the
  Coldest Brown Dwarfs}}.
\newblock \emph{\bibinfo{journal}{Astron. J.}} \textbf{\bibinfo{volume}{160}},
  \bibinfo{pages}{63} (\bibinfo{year}{2020}).

\bibitem{chabrier2014giant}
\bibinfo{author}{Chabrier, G.}, \bibinfo{author}{Johansen, A.},
  \bibinfo{author}{Janson, M.} \& \bibinfo{author}{Rafikov, R.}
\newblock \bibinfo{title}{Giant planet and brown dwarf formation}.
\newblock In \bibinfo{editor}{Beuther, H.}, \bibinfo{editor}{Klessen, R.~S.},
  \bibinfo{editor}{Dullemond, C.~P.} \& \bibinfo{editor}{Henning, T.} (eds.)
  \emph{\bibinfo{booktitle}{Protostars and Planets VI}},
  \bibinfo{pages}{619--642} (\bibinfo{publisher}{University of Arizona Press},
  \bibinfo{year}{2014}).

\bibitem{fletcher2014origin}
\bibinfo{author}{Fletcher, L.~N.} \emph{et~al.}
\newblock \bibinfo{title}{The origin of nitrogen on jupiter and saturn from the
  15n/14n ratio}.
\newblock \emph{\bibinfo{journal}{Icarus}} \textbf{\bibinfo{volume}{238}},
  \bibinfo{pages}{170--190} (\bibinfo{year}{2014}).

\bibitem{Oberg2019_Jupiter_Snowline}
\bibinfo{author}{{{\"O}berg}, K.~I.} \& \bibinfo{author}{{Wordsworth}, R.}
\newblock \bibinfo{title}{{Jupiter's Composition Suggests its Core Assembled
  Exterior to the N$_{2}$ Snowline}}.
\newblock \emph{\bibinfo{journal}{Astron. J.}} \textbf{\bibinfo{volume}{158}},
  \bibinfo{pages}{194} (\bibinfo{year}{2019}).

\bibitem{furuya2018depletion}
\bibinfo{author}{Furuya, K.} \& \bibinfo{author}{Aikawa, Y.}
\newblock \bibinfo{title}{Depletion of heavy nitrogen in the cold gas of
  star-forming regions}.
\newblock \emph{\bibinfo{journal}{Astrophys. J.}}
  \textbf{\bibinfo{volume}{857}}, \bibinfo{pages}{105} (\bibinfo{year}{2018}).

\bibitem{bergner2020evolutionary}
\bibinfo{author}{Bergner, J.~B.}, \bibinfo{author}{{\"O}berg, K.~I.},
  \bibinfo{author}{Bergin, E.~A.} \emph{et~al.}
\newblock \bibinfo{title}{An evolutionary study of volatile chemistry in
  protoplanetary disks}.
\newblock \emph{\bibinfo{journal}{Astrophys. J.}}
  \textbf{\bibinfo{volume}{898}}, \bibinfo{pages}{97} (\bibinfo{year}{2020}).

\bibitem{Visser2018_N_disks}
\bibinfo{author}{{Visser}, R.} \emph{et~al.}
\newblock \bibinfo{title}{{Nitrogen isotope fractionation in protoplanetary
  disks}}.
\newblock \emph{\bibinfo{journal}{Astron. Astrophys.}}
  \textbf{\bibinfo{volume}{615}}, \bibinfo{pages}{A75} (\bibinfo{year}{2018}).

\bibitem{bosman2019jupiter}
\bibinfo{author}{Bosman, A.~D.}, \bibinfo{author}{Cridland, A.~J.} \&
  \bibinfo{author}{Miguel, Y.}
\newblock \bibinfo{title}{Jupiter formed as a pebble pile around the n2 ice
  line}.
\newblock \emph{\bibinfo{journal}{Astron. Astrophys. Letters}}
  \textbf{\bibinfo{volume}{632}}, \bibinfo{pages}{L11} (\bibinfo{year}{2019}).

\bibitem{guillot2015}
\bibinfo{author}{{Guillot}, T.} \& \bibinfo{author}{{Gautier}, D.}
\newblock \bibinfo{title}{{Giant Planets}}.
\newblock In \bibinfo{editor}{{Schubert}, G.} (ed.)
  \emph{\bibinfo{booktitle}{Treatise on Geophysics}}, \bibinfo{pages}{529--557}
  (\bibinfo{year}{2015}).

\bibitem{suarez2022ultracool}
\bibinfo{author}{Suarez, G.} \& \bibinfo{author}{Metchev, S.}
\newblock \bibinfo{title}{Ultracool dwarfs observed with the spitzer infrared
  spectrograph. ii. emergence and sedimentation of silicate clouds in l dwarfs,
  and analysis of the full m5–t9 field dwarf spectroscopic sample}.
\newblock \emph{\bibinfo{journal}{Mon. Not. R. Astron. Soc.}}
  \textbf{\bibinfo{volume}{513}} (\bibinfo{year}{2022}).

\bibitem{Oberg_Bergin2021}
\bibinfo{author}{{{\"O}berg}, K.~I.} \& \bibinfo{author}{{Bergin}, E.~A.}
\newblock \bibinfo{title}{{Astrochemistry and compositions of planetary
  systems}}.
\newblock \emph{\bibinfo{journal}{Physics Reports}}
  \textbf{\bibinfo{volume}{893}}, \bibinfo{pages}{1--48}
  (\bibinfo{year}{2021}).

\bibitem{turrini_2021}
\bibinfo{author}{{Turrini}, D.} \emph{et~al.}
\newblock \bibinfo{title}{{Tracing the Formation History of Giant Planets in
  Protoplanetary Disks with Carbon, Oxygen, Nitrogen, and Sulfur}}.
\newblock \emph{\bibinfo{journal}{Astrophys. J.}}
  \textbf{\bibinfo{volume}{909}}, \bibinfo{pages}{40} (\bibinfo{year}{2021}).

\bibitem{Adams2021_CoreAccretion}
\bibinfo{author}{{Adams}, F.~C.}, \bibinfo{author}{{Meyer}, M.~R.} \&
  \bibinfo{author}{{Adams}, A.~D.}
\newblock \bibinfo{title}{{A Theoretical Framework for the Mass Distribution of
  Gas Giant Planets Forming through the Core Accretion Paradigm}}.
\newblock \emph{\bibinfo{journal}{Astrophys. J.}}
  \textbf{\bibinfo{volume}{909}}, \bibinfo{pages}{1} (\bibinfo{year}{2021}).

\bibitem{Marleau2019_Coreaccetion_HIP65426b}
\bibinfo{author}{{Marleau}, G.-D.}, \bibinfo{author}{{Coleman}, G. A.~L.},
  \bibinfo{author}{{Leleu}, A.} \& \bibinfo{author}{{Mordasini}, C.}
\newblock \bibinfo{title}{{Exploring the formation by core accretion and the
  luminosity evolution of directly imaged planets. The case of HIP 65426 b}}.
\newblock \emph{\bibinfo{journal}{Astron. Astrophys.}}
  \textbf{\bibinfo{volume}{624}}, \bibinfo{pages}{A20} (\bibinfo{year}{2019}).





{\bf References Methods}

  
\bibitem{Law2023}
\bibinfo{ }{{Law}, D.~R.} \emph{et~al.}
\newblock \bibinfo{title}{{A 3D Drizzle Algorithm for JWST and Practical
  Application to the MIRI Medium Resolution Spectrometer}}.
\newblock \emph{\bibinfo{journal}{arXiv e-prints}}
  \bibinfo{pages}{arXiv:2306.05520} (\bibinfo{year}{2023}).

\bibitem{Zalesky2019_RetrievalYdwarfs}
\bibinfo{ }{{Zalesky}, J.~A.}, \bibinfo{ }{{Line}, M.~R.},
  \bibinfo{ }{{Schneider}, A.~C.} \& \bibinfo{ }{{Patience}, J.}
\newblock \bibinfo{title}{{A Uniform Retrieval Analysis of Ultra-cool Dwarfs.
  III. Properties of Y Dwarfs}}.
\newblock \emph{\bibinfo{journal}{Astrophys. J.}}
  \textbf{\bibinfo{volume}{877}}, \bibinfo{pages}{24} (\bibinfo{year}{2019}).

\bibitem{morleyskemer2018}
\bibinfo{ }{{Morley}, C.~V.} \emph{et~al.}
\newblock \bibinfo{title}{{An L Band Spectrum of the Coldest Brown Dwarf}}.
\newblock \emph{\bibinfo{journal}{Astrophys. J.}}
  \textbf{\bibinfo{volume}{858}}, \bibinfo{pages}{97} (\bibinfo{year}{2018}).

\bibitem{manggao2022}
\bibinfo{ }{{Mang}, J.} \emph{et~al.}
\newblock \bibinfo{title}{{Microphysics of Water Clouds in the Atmospheres of Y
  Dwarfs and Temperate Giant Planets}}.
\newblock \emph{\bibinfo{journal}{Astrophys. J.}}
  \textbf{\bibinfo{volume}{927}}, \bibinfo{pages}{184} (\bibinfo{year}{2022}).

\bibitem{lineteske2015}
\bibinfo{ }{{Line}, M.~R.}, \bibinfo{ }{{Teske}, J.},
  \bibinfo{ }{{Burningham}, B.}, \bibinfo{ }{{Fortney}, J.~J.} \&
  \bibinfo{ }{{Marley}, M.~S.}
\newblock \bibinfo{title}{{Uniform Atmospheric Retrieval Analysis of Ultracool
  Dwarfs. I. Characterizing Benchmarks, Gl 570D and HD 3651B}}.
\newblock \emph{\bibinfo{journal}{Astrophys. J.}}
  \textbf{\bibinfo{volume}{807}}, \bibinfo{pages}{183} (\bibinfo{year}{2015}).

\bibitem{Tremblin2019}
\bibinfo{ }{{Tremblin}, P.} \emph{et~al.}
\newblock \bibinfo{title}{{Thermo-compositional Diabatic Convection in the
  Atmospheres of Brown Dwarfs and in Earth{\textquoteright}s Atmosphere and
  Oceans}}.
\newblock \emph{\bibinfo{journal}{Astrophys. J.}}
  \textbf{\bibinfo{volume}{876}}, \bibinfo{pages}{144} (\bibinfo{year}{2019}).

\bibitem{tsailyons2017}
\bibinfo{ }{{Tsai}, S.-M.} \emph{et~al.}
\newblock \bibinfo{title}{{VULCAN: An Open-source, Validated Chemical Kinetics
  Python Code for Exoplanetary Atmospheres}}.
\newblock \emph{\bibinfo{journal}{Astrophys. J. Sup. Series}}
  \textbf{\bibinfo{volume}{228}}, \bibinfo{pages}{20} (\bibinfo{year}{2017}).

\bibitem{Feroz2008sampling}
\bibinfo{ }{{Feroz}, F.} \& \bibinfo{ }{{Feroz}, M.~P.}
\newblock \bibinfo{title}{{Multimodal nested sampling: an efficient and robust
  alternative to Markov Chain Monte Carlo methods for astronomical data
  analyses}}.
\newblock \emph{\bibinfo{journal}{Mon. Not. R. Astron. Soc.}}
  \textbf{\bibinfo{volume}{384}}, \bibinfo{pages}{449--463}
  (\bibinfo{year}{2008}).

\bibitem{Buchner2014_BayesianSelection}
\bibinfo{ }{{Buchner}, J.} \emph{et~al.}
\newblock \bibinfo{title}{{X-ray spectral modelling of the AGN obscuring region
  in the CDFS: Bayesian model selection and catalogue}}.
\newblock \emph{\bibinfo{journal}{Astron. Astrophys.}}
  \textbf{\bibinfo{volume}{564}}, \bibinfo{pages}{A125} (\bibinfo{year}{2014}).

\bibitem{Polyansky2018_Exomol}
\bibinfo{ }{{Polyansky}, O.~L.} \emph{et~al.}
\newblock \bibinfo{title}{{ExoMol molecular line lists XXX: a complete
  high-accuracy line list for water}}.
\newblock \emph{\bibinfo{journal}{Mon. Not. R. Astron. Soc.}}
  \textbf{\bibinfo{volume}{480}}, \bibinfo{pages}{2597--2608}
  (\bibinfo{year}{2018}).

\bibitem{Hargreaves2020_Methane_HITEMP}
\bibinfo{ }{{Hargreaves}, R.~J.} \emph{et~al.}
\newblock \bibinfo{title}{{An Accurate, Extensive, and Practical Line List of
  Methane for the HITEMP Database}}.
\newblock \emph{\bibinfo{journal}{Astrophys. J. Sup. Series}}
  \textbf{\bibinfo{volume}{247}}, \bibinfo{pages}{55} (\bibinfo{year}{2020}).

\bibitem{rothman_2010}
\bibinfo{ }{{Rothman}, L.~S.} \emph{et~al.}
\newblock \bibinfo{title}{{HITEMP, the high-temperature molecular spectroscopic
  database}}.
\newblock \emph{\bibinfo{journal}{\jqsrt}} \textbf{\bibinfo{volume}{111}},
  \bibinfo{pages}{2139--2150} (\bibinfo{year}{2010}).

\bibitem{exo_co2}
\bibinfo{ }{Yurchenko, S.~N.}, \bibinfo{ }{Mellor, T.~M.},
  \bibinfo{ }{Freedman, R.~S.} \& \bibinfo{ }{Tennyson, J.}
\newblock \bibinfo{title}{{ExoMol line lists – XXXIX. Ro-vibrational
  molecular line list for CO2}}.
\newblock \emph{\bibinfo{journal}{Mon. Not. R. Astron. Soc.}}
  \textbf{\bibinfo{volume}{496}}, \bibinfo{pages}{5282--5291}
  (\bibinfo{year}{2020}).

\bibitem{rothman_2013}
\bibinfo{ }{{Rothman}, L.~S.} \emph{et~al.}
\newblock \bibinfo{title}{{The HITRAN2012 molecular spectroscopic database}}.
\newblock \emph{\bibinfo{journal}{\jqsrt}} \textbf{\bibinfo{volume}{130}},
  \bibinfo{pages}{4--50} (\bibinfo{year}{2013}).

\bibitem{exo_h2s}
\bibinfo{ }{Azzam, A. A.~A.}, \bibinfo{ }{Tennyson, J.},
  \bibinfo{ }{Yurchenko, S.~N.} \& \bibinfo{ }{Naumenko, O.~V.}
\newblock \bibinfo{title}{{ExoMol molecular line lists – XVI. The
  rotation–vibration spectrum of hot H2S}}.
\newblock \emph{\bibinfo{journal}{Mon. Not. R. Astron. Soc.}}
  \textbf{\bibinfo{volume}{460}}, \bibinfo{pages}{4063--4074}
  (\bibinfo{year}{2016}).

\bibitem{exo_nh3}
\bibinfo{ }{Coles, P.~A.}, \bibinfo{ }{Yurchenko, S.~N.} \&
  \bibinfo{ }{Tennyson, J.}
\newblock \bibinfo{title}{{ExoMol molecular line lists – XXXV. A
  rotation-vibration line list for hot ammonia}}.
\newblock \emph{\bibinfo{journal}{Mon. Not. R. Astron. Soc.}}
  \textbf{\bibinfo{volume}{490}}, \bibinfo{pages}{4638--4647}
  (\bibinfo{year}{2019}).

\bibitem{exo_ph3}
\bibinfo{ }{Sousa-Silva, C.}, \bibinfo{ }{Al-Refaie, A.~F.},
  \bibinfo{ }{Tennyson, J.} \& \bibinfo{ }{Yurchenko, S.~N.}
\newblock \bibinfo{title}{{ExoMol line lists – VII. The rotation–vibration
  spectrum of phosphine up to 1500 K}}.
\newblock \emph{\bibinfo{journal}{Mon. Not. R. Astron. Soc.}}
  \textbf{\bibinfo{volume}{446}}, \bibinfo{pages}{2337--2347}
  (\bibinfo{year}{2014}).

\bibitem{bennekeseager2013}
\bibinfo{ }{{Benneke}, B.} \& \bibinfo{ }{{Seager}, S.}
\newblock \bibinfo{title}{{How to Distinguish between Cloudy Mini-Neptunes and
  Water/Volatile-dominated Super-Earths}}.
\newblock \emph{\bibinfo{journal}{Astrophys. J.}}
  \textbf{\bibinfo{volume}{778}}, \bibinfo{pages}{153} (\bibinfo{year}{2013}).

\bibitem{ormelmin2019}
\bibinfo{ }{{Ormel}, C.~W.} \& \bibinfo{ }{{Min}, M.}
\newblock \bibinfo{title}{{ARCiS framework for exoplanet atmospheres. The cloud
  transport model}}.
\newblock \emph{\bibinfo{journal}{Astron. Astrophys.}}
  \textbf{\bibinfo{volume}{622}}, \bibinfo{pages}{A121} (\bibinfo{year}{2019}).

\bibitem{kawashimamin2021}
\bibinfo{ }{{Kawashima}, Y.} \& \bibinfo{ }{{Min}, M.}
\newblock \bibinfo{title}{{Implementation of disequilibrium chemistry to
  spectral retrieval code ARCiS and application to 16 exoplanet transmission
  spectra. Indication of disequilibrium chemistry for HD 209458b and
  WASP-39b}}.
\newblock \emph{\bibinfo{journal}{Astron. Astrophys.}}
  \textbf{\bibinfo{volume}{656}}, \bibinfo{pages}{A90} (\bibinfo{year}{2021}).

\bibitem{greenberg19a}
\bibinfo{ }{Greenberg, D.}, \bibinfo{ }{Nonnenmacher, M.} \&
  \bibinfo{ }{Macke, J.}
\newblock \bibinfo{title}{Automatic posterior transformation for
  likelihood-free inference}.
\newblock In \bibinfo{editor}{Chaudhuri, K.} \& \bibinfo{editor}{Salakhutdinov,
  R.} (eds.) \emph{\bibinfo{booktitle}{Proceedings of the 36th International
  Conference on Machine Learning}}, vol.~\bibinfo{volume}{97} of
  \emph{\bibinfo{series}{Proceedings of Machine Learning Research}},
  \bibinfo{pages}{2404--2414} (\bibinfo{publisher}{PMLR},
  \bibinfo{year}{2019}).

\bibitem{Burningham2017_Retrieval_CloudyL}
\bibinfo{ }{{Burningham}, B.} \emph{et~al.}
\newblock \bibinfo{title}{{Retrieval of atmospheric properties of cloudy L
  dwarfs}}.
\newblock \emph{\bibinfo{journal}{Mon. Not. R. Astron. Soc.}}
  \textbf{\bibinfo{volume}{470}}, \bibinfo{pages}{1177--1197}
  (\bibinfo{year}{2017}).

\bibitem{Burningham2021_Clouds_2M2244}
\bibinfo{ }{{Burningham}, B.} \emph{et~al.}
\newblock \bibinfo{title}{{Cloud busting: enstatite and quartz clouds in the
  atmosphere of 2M2224-0158}}.
\newblock \emph{\bibinfo{journal}{Mon. Not. R. Astron. Soc.}}
  \textbf{\bibinfo{volume}{506}}, \bibinfo{pages}{1944--1961}
  (\bibinfo{year}{2021}).

\bibitem{Toon1989_RadiativePhotodissociation}
\bibinfo{ }{{Toon}, O.~B.}, \bibinfo{ }{{McKay}, C.~P.},
  \bibinfo{ }{{Ackerman}, T.~P.} \& \bibinfo{ }{{Santhanam}, K.}
\newblock \bibinfo{title}{{Rapid calculation of radiative heating rates and
  photodissociation rates in inhomogeneous multiple scattering atmospheres}}.
\newblock \emph{\bibinfo{journal}{\jgr}} \textbf{\bibinfo{volume}{94}},
  \bibinfo{pages}{16287--16301} (\bibinfo{year}{1989}).

\bibitem{Madhusudhan2009_retrieval}
\bibinfo{ }{{Madhusudhan}, N.} \& \bibinfo{ }{{Seager}, S.}
\newblock \bibinfo{title}{{A Temperature and Abundance Retrieval Method for
  Exoplanet Atmospheres}}.
\newblock \emph{\bibinfo{journal}{Astrophys. J.}}
  \textbf{\bibinfo{volume}{707}}, \bibinfo{pages}{24--39}
  (\bibinfo{year}{2009}).

\bibitem{Freedman2008_Opacities}
\bibinfo{ }{{Freedman}, R.~S.}, \bibinfo{ }{{Marley}, M.~S.} \&
  \bibinfo{ }{{Lodders}, K.}
\newblock \bibinfo{title}{{Line and Mean Opacities for Ultracool Dwarfs and
  Extrasolar Planets}}.
\newblock \emph{\bibinfo{journal}{Astrophys. J. Sup. Series}}
  \textbf{\bibinfo{volume}{174}}, \bibinfo{pages}{504--513}
  (\bibinfo{year}{2008}).

\bibitem{Freedman2014_Opacities}
\bibinfo{ }{{Freedman}, R.~S.} \emph{et~al.}
\newblock \bibinfo{title}{{Gaseous Mean Opacities for Giant Planet and
  Ultracool Dwarf Atmospheres over a Range of Metallicities and Temperatures}}.
\newblock \emph{\bibinfo{journal}{Astrophys. J. Sup. Series}}
  \textbf{\bibinfo{volume}{214}}, \bibinfo{pages}{25} (\bibinfo{year}{2014}).

\bibitem{ForemanMackey2013_EMCEE}
\bibinfo{ }{{Foreman-Mackey}, D.}, \bibinfo{ }{{Hogg}, D.~W.},
  \bibinfo{ }{{Lang}, D.} \& \bibinfo{ }{{Goodman}, J.}
\newblock \bibinfo{title}{{emcee: The MCMC Hammer}}.
\newblock \emph{\bibinfo{journal}{\pasp}} \textbf{\bibinfo{volume}{125}},
  \bibinfo{pages}{306} (\bibinfo{year}{2013}).

\bibitem{schneiderbitsch2021}
\bibinfo{ }{{Schneider}, A.~D.} \& \bibinfo{ }{{Bitsch}, B.}
\newblock \bibinfo{title}{{How drifting and evaporating pebbles shape giant
  planets. II. Volatiles and refractories in atmospheres}}.
\newblock \emph{\bibinfo{journal}{Astron. Astrophys.}}
  \textbf{\bibinfo{volume}{654}}, \bibinfo{pages}{A72} (\bibinfo{year}{2021}).

\end{thebibliography}
\end{document}